\input harvmac
\input epsf
\def\figin{\epsfcheck\figin}\def\figins{\epsfcheck\figins}
\def\epsfcheck{\ifx\epsfbox\UnDeFiNeD
\message{(NO epsf.tex, FIGURES WILL BE IGNORED)}
\gdef\figin##1{\vskip2in}\gdef\figins##1{\hskip.5in}
\else\message{(FIGURES WILL BE INCLUDED)}%
\gdef\figin##1{##1}\gdef\figins##1{##1}\fi}
\def\DefWarn#1{}
\def\figinsert{\goodbreak\midinsert}
\def\ifig#1#2#3{\DefWarn#1\xdef#1{fig.~\the\figno}
\writedef{#1\leftbracket fig.\noexpand~\the\figno}%
\figinsert\figin{\centerline{#3}}\medskip\centerline{\vbox{\baselineskip12pt
\advance\hsize by -1truein\noindent\footnotefont{\bf Fig.~\the\figno:} #2}}
\bigskip\endinsert\global\advance\figno by1}

\def\l{\ell}
\def\hj{{\hat j}}
\def\hn{{\hat n}}
\def\hs{{\hat s}}

\def\p{\partial}

\def\mod{{\rm mod}}
\def\IR{\relax{\rm I\kern-.18em R}}
\def\IZ{\relax\ifmmode\hbox{Z\kern-.4em Z}\else{Z\kern-.4em Z}\fi}
\def\IF{\relax{\rm I\kern-.18em F}}
\def\IP{\relax{\rm I\kern-.18em P}}
\def\CS{{\cal S}}

\def\K3{{\bf K3}}
\def\journal#1&#2(#3){\unskip, \sl #1\ \bf #2 \rm(19#3) }
\def\andjournal#1&#2(#3){\sl #1~\bf #2 \rm (19#3) }

\def\bar{\overline}
\def\hat{\widehat}

\def\tilde{\widetilde}

\def\frac#1#2{{#1\over#2}}

\def\half{\frac12}

\def\inbar{\,\vrule height1.5ex width.4pt depth0pt}
\def\IC{\relax\hbox{$\inbar\kern-.3em{\rm C}$}}
\def\IR{\relax{\rm I\kern-.18em R}}
\def\IP{\relax{\rm I\kern-.18em P}}

%
%

%
\catcode`\@=11
\def\slash#1{\mathord{\mathpalette\c@ncel{#1}}}
\overfullrule=0pt

\def\CC{{\cal C}}

\def\NN{{\cal N}}

\def\underrel#1\over#2{\mathrel{\mathop{\kern\z@#1}\limits_{#2}}}

\catcode`\@=12


%

\def\mod{{\rm mod}}

\def\exp{{\rm exp}}



\lref\schrecent{
S.~Fredenhagen and V.~Schomerus,
``Brane dynamics in CFT backgrounds,''
hep-th/0104043. }

\lref\rogen{
A.~Recknagel, D.~Roggenkamp and V.~Schomerus,
``On relevant boundary perturbations of unitary minimal models,''
Nucl.\ Phys.\ B { 588}, 552 (2000)
[hep-th/0003110].
}

\lref\stanciu{
S.~Stanciu,
``D-branes in group manifolds,''
JHEP{ 0001}, 025 (2000)
[hep-th/9909163];
S.~Stanciu,
``D-branes in an AdS(3) background,''
JHEP{9909}, 028 (1999)
[hep-th/9901122].}

\lref\fstanciu{
J.~M.~Figueroa-O'Farrill and S.~Stanciu,
``D-branes in AdS(3) x S(3) x S(3) x S(1),''
JHEP{ 0004}, 005 (2000)
[hep-th/0001199].}

\lref\klimcik{C.~Klimcik and P.~Severa, ``Open strings and
D-branes in WZNW models,'' Nucl.\ Phys.\ B {\bf 488} (1997) 653
[hep-th/9609112].}

\lref\fuzzysph{
J.~Madore,
Class.\ Quant.\ Grav.\ {\bf 9}, 69 (1992)
}

\lref\hikida{ Y.~Hikida, M.~Nozaki and Y.~Sugawara, ``Formation
of spherical D2-brane from multiple D0-branes,'' hep-th/0101211. }

\lref\buscher{
T.~H.~Buscher,
``Path Integral Derivation Of Quantum Duality In Nonlinear Sigma Models,''
Phys.\ Lett.\ B { 201}, 466 (1988).
}

\lref\ars{A. Alekseev, A. Recknagel, and V. Schomerus,
``Non-commutative World-volume Geometries: Branes on SU(2) and
Fuzzy Spheres,'' hep-th/9908040; ``Brane Dynamics in Background
Fluxes and Non-commutative Geometry,'' hep-th/0003187; ``Open
Strings and Non-commutative Geometry of Branes on Group
Manifolds,'' hep-th/0104054. }

\lref\as{A. Alekseev and V. Schomerus,
 ``RR charges of D2-branes in the WZW model,''
hep-th/0007090;
``D-branes in the WZW model,''
Phys.\ Rev.\ D { 60}, 061901 (1999)
[hep-th/9812193].
}
 \lref\ooy{H.~Ooguri, Y.~Oz and Z.~Yin,
``D-branes on Calabi-Yau spaces and their mirrors,'' Nucl.\
Phys.\ B {477} (1996) 407 [hep-th/9606112].}

\lref\bds{C. Bachas, M. Douglas, and C. Schweigert,
``Flux Stabilization of D-branes,''hep-th/0003037 }
\lref\behrend{R.E.  Behrend, P.A. Pearce, V.B. Petkova,
J-B Zuber, ``Boundary conditions in Rational
Conformal Field Theories,'' hep-th/9908036}
\lref\bfk{W. Boucher, D. Friedan, and A. Kent,
``Determinant formulae and unitarity for the
N=2 superconformal algebras in two dimensions
or exact results on string compactification,''
Phys. Lett. { 172B}(1986)316}

\lref\bouwknegt{P. Bouwknegt and V. Mathai, ``D-Branes,
B-fields, and Twisted K-Theory,'' hep-th/0002023.}
\lref\bdlr{I. Brunner, M.R. Douglas, A. Lawrence,
and C. Romelsberger,  ``D-branes on the Quintic,''
hep-th/9906200}
\lref\brunnerdistler{I. Brunner and J. Distler,
``Torsion D-Branes in Nongeometrical Phases,''
hep-th/0102018}
\lref\brunnermoore{I. Brunner and G. Moore,
unpublished.}

\lref\cardy{J. Cardy, ``Boundary conditions, fusion
rules and the Verlinde formula,''
Nucl. Phys. B324 (1989)581}

\lref\cmnp{A. Cohen, G. Moore, P. Nelson, and J. Polchinski,
``An invariant string propagator,'' in {\it Unified String
Theories}, p. 568,  M. Green and D. Gross, eds. World Scientific,
1986. }
\lref\falceto{F. Falceto and K. Gawedzki,
``Chern-Simons States at Genus One,''
Commun. Math. Phys. 159 (1994) 549-580.}
\lref\fffs{G. Felder, J. Fr\"ohlich, J. Fuchs, C. Schweigert,
``The geometry of WZW branes,''hep-th/9909030, J.Geom.Phys. 34
(2000) 162-190}
\lref\fredenhagen{S. Fredenhagen and V. Schomerus,
``Branes on Group Manifolds, Gluon Condensates, and twisted K-theory,''
hep-th/0012164}
\lref\fht{D. Freed,
``The Verlinde algebra is twisted equivariant K-theory,''
math.RT/0101038}

\lref\fss{J. Fuchs, A.N. Schellekens, and C. Schweigert,
``A matrix S for all simple current extensions,''
hep-th/9601078; Nucl.Phys. B473 (1996) 3.}

 \lref\fs{J. Fuchs and C.
Schweigert, ``A classifying algebra for boundary conditions,''
hep-th/9708141, Phys.Lett. B414 (1997) 251-259 \semi
 ``Branes: from free fields to general
backgrounds,'' Nucl. Phys. { B530} (1998) 99,
hep-th/9712257\semi  ``Symmetry breaking boundaries I. General
theory,'' hep-th/9902132\semi ``Symmetry breaking boundaries. II:
More structures, examples,'' Nucl.\ Phys.\ B { 568} (2000) 543
[hep-th/9908025].}

\lref\gepner{D. Gepner, ``Space-time supersymmetry in
compactified string theory and superconformal models,'' Nucl.
Phys. { B296}(1988)757}

\lref\gepnerqiu{D. Gepner and Z. Qiu, Nucl. Phys. { B285}(1987)423}

\lref\tensiondimension{J.A. Harvey, S. Kachru,
G. Moore, and E. Silverstein, ``Tension is
Dimension,'' hep-th/9909072}

\lref\HIV{K. Hori, A. Iqbal, and C. Vafa, ``D-Branes and Mirror
Symmetry,'' hep-th/0005247}

\lref\hori{K. Hori, ``Linear Models of Supersymmetric
D-Branes,'' hep-th/0012179}

\lref\ishibashi{N. Ishibashi, ``The Boundary and Crosscap States
in Conformal Field Theories'', {\it Mod. Phys. Lett.} { A4}
(1989) 251; N. Ishibashi and T. Onagi, ``Conformal Field Theories
on Boundaries and Surfaces with Crosscaps'', {\it Mod. Phys.
Lett.} { A4} (1989) 161.}

\lref\lazaroiu{C.I. Lazaroiu, `` Unitarity, D-brane dynamics and
D-brane categories,'' hep-th/0102183; ``Generalized complexes and
string field theory,'' hep-th/0102122; ``Instanton amplitudes in
open-closed topological string theory,'' hep-th/0011257; ``On the
structure of open-closed topological field theory in two
dimensions,'' hep-th/0010269.}

\lref\lewellen{D. Lewellen, ``Sewing constraints
for conformal field theories on surfaces with
boundaries,'' Nucl.Phys. { B372} (1992) 654;
J.L. Cardy and D.C. Lewellen, ``Bulk and boundary
operators in conformal field theory,''
Phys.Lett. { B259} (1991) 274.}

\lref\runkel{ I.~Runkel, ``Boundary structure constants for the
A-series Virasoro minimal  models,'' Nucl.\ Phys.\ B { 549}, 563
(1999) [hep-th/9811178]. }

\lref\mooresegal{G. Moore and G. Segal, to appear, sometime.}

\lref\mooreseiberg{ G. Moore and N. Seiberg,
``Classical and Quantum Conformal Field Theory,''
Commun. Math. Phys. {\bf 123}(1989)177;
``Lectures on Rational Conformal Field Theory,''
in {\it Strings '89},Proceedings
of the Trieste Spring School on Superstrings,
3-14 April 1989, M. Green, et. al. Eds. World
Scientific, 1990}

\lref\bfs{ L.~Birke, J.~Fuchs and C.~Schweigert, ``Symmetry
breaking boundary conditions and WZW orbifolds,'' Adv.\ Theor.\
Math.\ Phys.\ { 3}, 671 (1999) [hep-th/9905038].}

\lref\jp{J. Polchinski, {\it String Theory}, Cambridge Univ.
Press, 1998}

\lref\qiu{Z. Qiu, ``Modular invariant partition
functions for N=2 superconformal field theories,''
Phys. Lett. { B198}(1987)497}

\lref\ravanini{F. Ravanini and S-K. Yang,
``Modular invariance in N=2 superconformal field
theories,'' Phys. Lett. { B195}(1987)202}

\lref\reck{A. Recknagel and
V. Schomerus, ``D-branes in Gepner models,'' Nucl. Phys. {
B531} (1998) 185; hep-th/9712186.}

 \lref\sagnotti{G.
Pradisi, A. Sagnotii, and Ya. S. Stanev, ``Completeness conditions
for boundary operators in 2D conformal field theory,'' Phys. Lett.
{ B381} (1996) 97; hep-th/9603097.}

\lref\schwimmerseiberg{A. Schwimmer and N. Seiberg,
Phys. Lett. { 184B} (1987) 191}

\lref\sw{N. Seiberg and E. Witten, ``String Theory and Noncommutative
Geometry,'' JHEP { 9909}:032,1999; hep-th/9908142.}

\lref\disksref{
V.~Sahakian, ``On D0 brane polarization by tidal forces,''
hep-th/0102200.
}

\lref\gps{ S.~B.~Giddings, J.~Polchinski and A.~Strominger,
``Four-dimensional black holes in string theory,'' Phys.\ Rev.\ D
{ 48}, 5784 (1993) [hep-th/9305083]. } \lref\myers{R.~C.~Myers,
``Dielectric-branes,'' JHEP{ 9912}, 022 (1999) [hep-th/9910053]. }
\lref\bachas{ C.~Bachas and M.~Petropoulos, ``Anti-de-Sitter
D-branes,'' JHEP{ 0102}, 025 (2001) [hep-th/0012234]. }

\lref\hirosi{H. Ooguri, talk at David Gross birthday conference.
}

\lref\zam{ N.~Seiberg, ``Notes On Quantum Liouville Theory And
Quantum Gravity,'' Prog.\ Theor.\ Phys.\ Suppl.\ {\bf 102} (1990)
319; G.~Moore, N.~Seiberg and M.~Staudacher, ``From loops to
states in 2-D quantum gravity,'' Nucl.\ Phys.\ B {\bf 362} (1991)
665; G.~Moore and N.~Seiberg, ``From loops to fields in 2-D
quantum gravity,'' Int.\ J.\ Mod.\ Phys.\ A {\bf 7} (1992) 2601;
A.~Zamolodchikov and A.~Zamolodchikov, ``Liouville field theory
on a pseudosphere,'' hep-th/0101152. }

\lref\suss{
J.~McGreevy, L.~Susskind and N.~Toumbas,
``Invasion of the giant gravitons from anti-de Sitter space,''
JHEP{ 0006}, 008 (2000)
[hep-th/0003075].
}

\lref\arched{ S.~Chakravarty, K.~Dasgupta, O.~J.~Ganor and
G.~Rajesh, ``Pinned branes and new non Lorentz invariant
theories,'' Nucl.\ Phys.\ B { 587}, 228 (2000) [hep-th/0002175];
K.~Dasgupta, O.~J.~Ganor and G.~Rajesh, ``Vector deformations of
N = 4 super-Yang-Mills theory, pinned branes, and arched
strings,'' hep-th/0010072 . }

\lref\wittengg{E. Witten,
``The Verlinde Algebra And The Cohomology Of The Grassmannian,''
 hep-th/9312104. }

\Title{\vbox{\baselineskip12pt
\hbox{hep-th/0105038}
\hbox{RUNHETC-2001-15 }}}%
{\vbox{\centerline{Geometrical interpretation of D-branes }
\medskip
\centerline{in gauged  WZW models}
 }}

\smallskip
\centerline{Juan Maldacena$^{1,2}$, Gregory Moore$^{3}$, Nathan
Seiberg$^{1}$ }
\medskip

\centerline{\it $^{1}$ School of Natural Sciences,}
\centerline{\it Institute for Advanced Study} \centerline{\it
Einstein Drive} \centerline{\it Princeton, New Jersey, 08540}

\centerline{\it $^{2}$ Department of Physics, Harvard University}
\centerline{\it Cambridge, MA 02138}

\centerline{\it $^{3}$ Department of Physics, Rutgers University}
\centerline{\it Piscataway, New Jersey, 08855-0849}

\smallskip
\noindent
 We show that one can construct D-branes in parafermionic
and WZW theories (and their orbifolds) which have very natural
geometrical interpretations, and yet are not automatically
included in the standard Cardy construction of D-branes in
rational conformal field theory. The relation between these
theories and their T-dual description leads to an analogy between
these D-branes and the familiar A-branes and B-branes of $N=2$
theories.

\vglue .3cm

\bigskip
\noindent

\Date{May 3, 2001}


\newsec{Introduction}

D-branes are an essential element of modern string theory \jp.
Therefore, given an on-shell closed string background a
fundamental question is: ``What are the D-branes in the given
background?''  This innocent question turns out to be a difficult
and subtle problem. A useful approach in answering this question
is to phrase it in the language of boundary states in abstract
conformal field theory. General sewing conditions for open string
boundary conditions are well-understood mostly in rational
conformal field theory - RCFT (for a partial list of references
see, e.g.\ \refs{\cardy\lewellen\runkel\sagnotti
\fs\bfs\reck\lazaroiu-\mooresegal}). Moreover, it seems clear
that one should also impose local, conformally invariant  boundary
conditions. Combining these conditions in a useful and general
way is an outstanding unsolved problem.

With this motivation in mind (as well as others) many authors
have studied D-branes in the solvable WZW models for compact
Lie group $G$ using the exact solution of the conformal field
theory. A beautiful picture has begun to emerge
in which, among other things,  D0-branes blow up into
branes wrapping conjugacy classes in the group
\refs{\ars\bds\as-\fredenhagen}. The next interesting case to
study is that of gauged WZW models, and the present paper
addresses the geometrical interpretation of branes in
those models in the case where $G$ is $SU(2)$.

The subgroups of $G=SU(2)$ which we can gauge include the $ADE$
discrete subgroups, the $U(1)$ subgroup, and $G$ itself. These
groups can act on the left or the right. The $G/G$ model is a
topological theory which has been studied, for example, in
\wittengg.  As noted in \brunnermoore\  this model is the basis
for a physical explanation of  the recent mathematical result of
\fht. Indeed, the branes very naturally correspond to a basis for
the Verlinde algebra. Other gaugings will be the focus of the
present paper.  Indeed, we will focus on the $SU(2)/U(1)$ model,
where we gauge a vector (or axial) $U(1)$ and on the
$SU(2)/\IZ_n$ models where we gauge a left action of $\IZ_n$. The
generalization to other discrete subgroup actions is left as an
interesting open problem.

The novelty of the results below is that even in these
well-studied RCFT's it is possible to go beyond the D-branes
constructed via the standard Cardy theory while still maintaining
a good geometrical picture. This is not to say that the branes
were completely unknown. A general theory of symmetry breaking
branes was set forth  in \refs{\fs,\bfs} and we believe the
branes constructed below fit into that theory. Moreover, D-branes
in parafermionic models were studied in \reck\ from an algebraic
perspective. The emphasis in the present paper is on the {\it
geometrical} interpretation of the theory.

Geometrically, the $SU(2)/U(1)$ parafermion theory is a sigma
model with a disk target space.  In this paper we show that  the
simple D-branes are D0-branes at special points at the boundary
of the disk together with  D1-branes stretched between these
points. One of the interesting aspects of these D-branes is the
interplay between the exact algebraic description of the branes
and their geometric description in the target space.

In bosonic theories the D-branes are not oriented.  In
supersymmetric theories the D-branes can be oriented.  In the
$SU(2)$ theory the D0 and D2-branes become orientable and they
each have a moduli space $S^3$ corresponding to translations
around the group.  The supersymmetric $SU(2)/U(1)$ theory has
D1-branes and D2-branes.  Some of them are orientable.

One of our main results is that in addition to these branes which
we refer to as A-branes these theories also have other branes. We
refer to the new branes as B-branes.  This terminology is
borrowed from $\NN=2$ theories \ooy\ and is being extended here.
The B-branes can be found by the following procedure. These
theories are T-dual to their $\IZ_k$ orbifolds. The A-branes in
the covering theory lead, using the method of images, to some
branes in the orbifold theory.  The T-dual version of these brane
are not the original A-branes in the model we started with. These
are our new B-branes. The construction seems to fit into the
general scheme proposed in \refs{\fs,\bfs} for analyzing symmetry
breaking boundary conditions.

The B-branes in the $SU(2)$ theory are like D1-branes wrapping
around the group.  Several such D1-branes blow up into a
D3-brane. The moduli space of these branes is $SU(2)\times
SU(2)\times U(1)/U(1)^2$.

The B-branes in the parafermion theory are a D0-brane and
D2-branes at the center of the disk and most of them are
unstable.  Of particular interest is the case of $k$ even where a
bound state of ${k\over 2}+1$ such D0-branes at the center of the
disk is two separate D2-branes. Each of them covers the whole
disk and is stable.

In section 2 we review some well known results about conformal
field theories whose chiral algebras are based on $U(1)$ and
$SU(2)$ affine Lie algebras and the GKO $SU(2)/U(1)$ coset
theory.  Here we discuss only the closed strings.  In section 3
we start by considering the D-branes in the $U(1)$ theory.  This
elementary example is useful since it demonstrates our method in
a widely known context. We then turn to the D-branes in the
parafermion theory.  We discuss the A-branes and the B-branes and
study their properties.  In particular, we compute the spectrum
of open strings living on the various branes or stretched between
two different branes. In section 4 we present two effective
descriptions of these B-branes.  The first is as a bound state of
D0-branes (as in \refs{\ars,\disksref}) and the second is an
effective theory of a D2-brane, as in \bds . Section 5 is devoted
to the superparafermion theory and its D-branes and builds on the
results of \reck .  The A-branes are found to be oriented
D1-branes.  Most of the B-branes are unoriented D2-branes but a
few of them are oriented D2-branes. In section 6 we consider the
$SU(2)_k$ theory and its $\IZ_{k_1}$ orbifolds (Lens spaces).  In
addition to the known A-branes we also discuss the new B-branes
and explore some of their properties.

In several appendices we discuss some more technical details.  In
appendix A we list the $U(1)$ and $SU(2)$ characters and their
modular properties.  In appendix B we compute the open string
spectrum between A-branes and B-branes in the parafermion
theory.  Appendix C is devoted to three explicit examples of
parafermion theories which exhibit special features.  In appendix
D we present a calculation of the shape of the various branes
which substantiates our geometrical pictures.  Appendix E is a
review of $\NN=2$ theories and appendix F is an explicit analysis
of the first superparafermion theory.

In addition to the intrinsic interest in these models as solvable
conformal field theories, we would like to mention a few other
applications of these theories and their D-branes:
\item{1.}  The $SU(2)$ theory arises in string theory in the
background of NS5-branes and in condensed matter theory in the
Kondo problem as well as in other places.
\item{2.}  The $SU(2)$ theory is closely related to $SL(2)$ which
appears in string theory in the contexts of two and three
dimensional black holes.
\item{3.}  The $SU(2)_k/\IZ_{k_1}$ theory arises when $k/k_1$
NS5-branes and $k_1$ Kaluza-Klein monopoles coincide.
\item{4.}  Parafermions appear in many string constructions.  In
particular, they are closely related to the two dimensional black
hole, both the Lorentzian and Euclidean (cigar) versions.
\item{5.}  The $\NN=2$ super-discrete series is used as a building
block in Calabi-Yau compactifications. Some of the B-branes
constructed in the $\NN=2$ minimal models  might be very
interesting in understanding further the results of Brunner and
Distler \brunnerdistler\ on torsion branes.

\noindent
 We hope that our techniques will be useful in the study of these
models.

\newsec{Review of closed strings in parafermion theory}

\subsec{Algebraic considerations}

\centerline{$U(1)_k$}

The $U(1)_k$ chiral algebra with $k\in \IZ$ extends the chiral
algebra generated by a Gaussian $U(1)$ current $J=i\sqrt{2k} \p
X$ by including charged fields  of dimension $k$ and  charge $\pm
2k$,  $\exp[\pm i \sqrt{2k}X(z)]$.   The representations of
$U(1)_k$ are labeled by an integer $n$, defined modulo $2k$, and
identified with the charge.   The lowest $L_0$ eigenvalue in the
representation labeled by $n$ is
 \eqn\lowuo{\Delta_n={n^2\over 4k}}
 where we choose the fundamental domain $n=-k+1,-k+2,...,k$. The
character of this representation is
\eqn\character{
{\Tr}_{\CH_n} q^{L_0-1/24} e^{2\pi i z J_0} =
{\Theta_{n,k}(\tau,2z)\over \eta(\tau) }
}
(See appendix A for conventions on theta functions.)
We often specialize to $z=0$ and define
\eqn\uonecha{\psi_n (q)={1\over \eta(q)}\sum_{r\in \IZ + {n\over
2k}} q^{kr^2} .}
 The fusion rules are $n_1 \times n_2= (n_1+n_2 )\mod 2k $. The
action of the modular transformation $S$ on the characters is
\eqn\uonetra{\psi_n(q')={1\over \sqrt {2k}} \sum_{n'} e^{-i\pi nn'
\over k} \psi_{n'}(q) \qquad q=e^{2\pi i \tau} \qquad \tau'=-{1\over
\tau}.}

The diagonal modular invariant made of these representations is
 \eqn\diauo{Z_{diag}=\sum_n|\psi_n(q)|^2}
 and it describes a boson on a circle of radius $\sqrt{2k}$ (in units
where $\alpha'=2$).
The most general modular invariant is
 \eqn\nondiauo{Z_l=\sum_{n+\bar n=0\mod 2l\atop n-\bar n =0\mod 2l'}
\psi_n(q)\psi_{\bar n}(\bar q) ~~~~~~~~~ll'=k, ~~~~~~~l,l'\in
\IZ.}
 Clearly, this partition function describes a model which is a
$\IZ_l$ orbifold of the original model \diauo; i.e.\ a boson on a
circle of radius ${\sqrt{2k}\over l}=\sqrt{2l'\over l}$.  Since
$\psi_n=\psi_{-n}$, we have $Z_l=Z_{l'}$.  Of course, this is a
special case of T-duality.

\centerline{$SU(2)_k$}

The irreducible representations are labeled by $j=0,\half,...,{k
\over 2}$. The lowest $L_0$ eigenvalue in the $j^{th}$
representation is $\Delta_j={j(j+1) \over k+2}$.  The characters
$\chi_j (q)$ transform under the nontrivial modular
transformation $S$ as
\eqn\suttra{\chi_j(q')=\sum_{j'}S_{j}^{~j'}\chi_{j'}(q) \qquad
S_{j}^{~j'}=\sqrt{2\over k+2} \sin{\pi (2j+1)(2j'+1) \over k+2}.}
The fusion rules are
\eqn\sutwofr{ N_{jj'}^{~~j''} = \cases{
 1 & $\vert j-j'\vert\leq j'' \leq \min\{j+j',k -j-j'\}$ \cr
& \& $ j+j'+j'' \in \IZ $\cr
 0 & {\rm otherwise}
\cr} }

\centerline{Parafermions $\CA^{PF(k)}= {SU(2)_k\over U(1)_k}$}

The chiral algebra $\CA^{PF(k)}$ of this theory has a set of
irreducible representations $\CH_{(j,n)}$ described by pairs
$(j,n)$ where $j\in \half \IZ$, $0 \leq j \leq k/2$, and $n$ is
an integer defined modulo $2k$. The pairs are subject to a
constraint $2j+n=0 \mod 2$, and an equivalence relation $(j,n)
\sim ({k\over 2}-j,k+n)$. We will denote the set of distinct
irreducible representations by $PF(k) = \{(j,n)\}$. The character
of the representation $(j,n)$, denoted $\chi_{jn}(q)$, is
determined implicitly  by the decomposition
\eqn\parafdeco{\chi_j^{SU(2)}(\tau,z)=\sum_{n=-k}^{k+1}
\chi_{j,n}^{(k)}(q){\Theta_{n,k}(\tau,2z)\over \eta(\tau)} , }
where $q= \exp[2\pi i \tau]$.  Explicitly we have \gepnerqiu
\eqn\PFchar{
\eqalign{
\chi_{(j,n)}^{(k)}(\tau) & := {\Tr}_{\CH_{(j,n)}} q^{L_0-c/24} \cr
& = {1\over \eta(\tau)^2} \biggl(\sum_{(x,y)\in A(j,n,k)}+
\sum_{(x,y)\in A({k\over 2} -j, k+n,k)} \biggr) {\rm sign}(x)
e^{2\pi i \tau [(k+2)x^2-k y^2]} \cr}
}
where
\eqn\definofa{
A(j,n,k)=\left\{ (x,y): -\vert x \vert \leq y \leq \vert x\vert, \quad
 (x,y)\in \IZ^2 + (j+{1 \over 2(k+2)},{n\over 2k}) \right\} .}

The characters $\chi_{(j,n)}^{(k)}(\tau )$ are defined to be zero
if  $2j+n=1\mod 2$. We often abbreviate the notation to
$\chi_{jn}$, and the characters satisfy
\eqn\pararel{\chi_{j,n}=\chi_{j,-n}=\chi_{{k\over 2}-j,k-n}.}
(Warning: The representations labeled by $(j,n)$ and by $(j,-n)$
are distinct, but $(j,n)$ and $({k\over 2} - j, k + n)$ are the
same.) The lowest $L_0$ eigenvalue in the $(j,n)$ representation
is \eqn\lowestlz{\Delta_{jn}=\cases{{j(j+1)\over k+2} -{n^2\over
4k} &$-j\le {n\over 2} \le j $\cr {j(j+1)\over k+2} -{n^2\over
4k}+{n-2j\over 2}
 & $j\le {n\over 2} \le k-j$ \cr}}
where we shifted the fundamental domain of $n$ to be
$n=-2j,-2j+2,...,2k-2j-2$ which automatically implements the
selection rule $n+2j\in 2\IZ$.  Using \uonetra\suttra\ the action
of the modular group on the characters is
\eqn\charatrans{\chi_{jn}(q')=
\sum_{(j',n')\in PF(k) } {S^{PF}_{(j,n)}}^{~~(j',n')}
\chi_{j'n'}(q),}
where $q'=\exp[-2\pi i /\tau]$,
and the PF S-matrix is
\eqn\PFessmatrx{
{S^{PF}_{(j,n)}}^{~~(j',n')}= {\sqrt {2\over k}} e^{i\pi nn'
\over k}S_{j}^{~j'}. }
The fusion rules of the theory are
\eqn\PFfrs{
N_{(jn),(j'n')}^{~~~~~~~(j''n'')} := N_{jj'}^{~~j''}
\delta_{n+n'-n''}
+ N_{jj'}^{~~{k\over 2}-j''} \delta_{n+n'-n''-k}
}
where $\delta$ is a periodic delta function modulo $2k$.

When combining left and right-movers, the simplest modular
invariant partition function of the parafermion theory is
obtained by summing over all distinct representations
 \eqn\paraparti{Z=\sum_{(j,n)\in PF(k)}|\chi_{jn}|^2 =
\sum_j \sum_{n=1}^{2k} \half |\chi_{jn}|^2 ~ .}

In the semiclassical limit as $k$ goes to infinity with fixed $n$
and $j$ only the primary states of the parafermion algebra
survive.  Also, from \lowestlz\ it is clear that the spectrum
includes states with every value of $j=0,\half,1,...$ and for
each $j$ the states are labeled by the integer $n$ satisfying
$-j\le {n\over 2} \le j $ and $2j+n \in 2\IZ$.  Their dimension
is $\Delta_{jn}\approx {1\over k} \left [j(j+1) -{n^2\over
4}\right]$.

\subsec{Sigma model description}

One of the main themes of  this paper is that there are very
natural and intuitive geometrical descriptions of the intricate
algebraic formulae of RCFT. Accordingly, let us give the sigma
model description of the results of the previous section.

We begin with the $SU(2)$ level $k$ WZW model.
We can write the metric on $S^3$ in the following ways,
all of which will be useful.
\eqn\metricsth{\eqalign{
ds^2 &=  d\theta^2 + \sin^2 \theta d { \phi}^2 + \cos^2 \theta
d \tilde \phi^2
\cr
ds^2 & = {1 \over 4} \left[ (d \chi + \cos\tilde\theta d \varphi )^2
+ d {\tilde \theta}^2 + \sin^2{\tilde \theta} d \varphi^2 \right]
\cr
ds^2 & = d \psi^2 + \sin^2 \psi ds^2_{S^2}
\cr
 & ~~~~ \chi = \tilde \phi +  \phi ~,~~~~~ \varphi = \tilde \phi -
 \phi ~,~~~~~
\tilde \theta  = 2 \theta
\cr
& ~~~ g = e^{ i \chi { \sigma^3 \over 2}} e^{ i \tilde \theta
 { \sigma^1 \over 2}}
 e^{ i \varphi { \sigma^3 \over 2}} = e^{ i 2 \psi { \vec n . \vec
\sigma \over 2}}
}}
 where the last line denotes the $SU(2)$ group element in terms of
the Euler angles. The ranges are $0\leq \tilde\theta\leq \pi,
0\leq \varphi , \phi, \tilde \phi \leq 2\pi, 0\leq \chi\leq 4 \pi$
while $0\leq \psi\leq \pi$. In the first parametrization the
sphere is presented as $|z_1|^2+|z_2|^2=1$ with
$z_1=e^{i\phi}\sin \theta$; $z_2=e^{i\tilde \phi} \cos \theta$.
Notice also that the angle $\psi$ is related to a rotation by
angle $2 \psi$ and $\vec n$ is a unit vector (a point on $S^2$).

The Lagrangian for the $SU(2)$  model  can be written  as
\eqn\sigmamod{\eqalign{
S = & k \int \partial \theta \bar \partial \theta + \sin^2 \theta
\partial  \phi
 \bar \partial  \phi + \cos^2 \theta
\partial  \tilde \phi
 \bar \partial  \tilde \phi  +
\sin^2 \theta ( \partial  \phi \bar \partial \tilde  \phi
-  \bar \partial  \phi \partial \tilde  \phi )
\cr
 = & k\int
\biggl[ \partial \theta \bar \partial \theta + \tan^2 \theta
\partial  \phi
 \bar \partial  \phi + \cos^2 \theta
( \partial  \tilde \phi + \tan^2\theta \partial  \phi )
( \bar \partial \tilde  \phi -
\tan^2\theta \bar \partial  \phi)\biggr]
}}
 In the second line we essentially completed a square.
Geometrically, the GKO coset amounts to gauging a
$U(1)$ symmetry of the model corresponding to shifting $\tilde
\phi $. This is achieved by adding a gauge field to the last term
in the second line of \sigmamod\ so that we get
 \eqn\gaugedwzw{ S
= k\int \biggl[ \partial \theta \bar \partial \theta + \tan^2
\theta
\partial  \phi
 \bar \partial  \phi + \cos^2 \theta
( \partial  \tilde \phi + \tan^2\theta \partial  \phi + A_z)
( \bar \partial \tilde  \phi -
\tan^2\theta \bar \partial  \phi + A_{\bar z})\biggr]
}
Integrating out $A$ removes completely the last term and
produces a dilaton proportional to $\log \cos \theta $.

The net result is that the level $k$ parafermion theory is
described by a sigma model with metric and string coupling
 \eqn\parametric{\eqalign{
 &(ds)^2=k{1\over 1-\rho^2}(d\rho^2+\rho^2d\phi^2)\cr
 &g_s(\rho)\equiv e^{\Phi} =g_s(0)(1-\rho^2)^{-{1\over 2}}.}}
The relation between the coordinates in the metric \sigmamod\ and
the metric \parametric\ is $\rho = \sin \theta$. The topology of
the target space is a disk. Geometrically it has finite radial
geodesic distance,  but infinite circumference. There is a
curvature singularity at $\rho=1$. From \parametric\ it would
appear that the sigma model has  a $U(1)$ symmetry of shifts of
$\phi$. In fact the symmetry is broken to $\IZ_k$. More
precisely, we can check from \gaugedwzw\ that the current
corresponding to shifts in $\phi$ has a divergence proportional
to $\partial_\alpha j^\alpha \sim  k  F_{z\bar z}$ where $F$ is
the field strength of the gauge field. If we integrate this we
see that angular momentum in the $\phi$ direction is violated by
$k$ units. This is the familiar statement that if we gauge the
vector $U(1)$, then the axial $U(1)$ is anomalous and vice versa.
A $\IZ_k$ subgroup of $U(1)$ is non-anomalous and is a good
symmetry of the theory.

We can also relate this sigma model description to the usual
algebraic description of the coset. First notice that $J^3_L$ and
$J^3_R$ correspond to translations in the angles $\chi, -
\varphi$ respectively (see \metricsth ). In the parafermion
theory we gauge $J^3_L - J^3_R$, which corresponds to
translations in $\tilde \phi$. So starting from an $SU(2)$ state
$| \Psi\rangle $,  we impose the constraints $J^3_{L,n}
|\Psi\rangle = J^3_{R,n} |\Psi\rangle =0$, for $n>0$ and {\it
also} $(J^3_{L,0} - J^3_{R,0} ) |\Psi\rangle =0$. This is an
explicit way to parametrize the states in the coset.

For large $k$ we can analyze the spectrum of low dimension operators
by considering the ``space-time'' Lagrangian
\eqn\psilagclo{\sqrt g e^{-2\Phi}{{1\over 2} g^{ab}\partial_a\Psi
\partial_b\Psi}}
 which depends on the field $\Psi$, and the metric and dilaton
\parametric.  $\Psi$ is the wave function of a massless field that
has no indices in the directions of the parafermion space.  The
small fluctuations of $\Psi$ are controlled by the eigenvalue
equation
\eqn\psieigen{[-{1\over
2}\nabla^2 + \nabla\Phi\nabla - 2\Delta]\Psi=0}
(the factor of $2$ in front of $\Delta$ is because the eigenvalue is
$\Delta+\overline \Delta=2\Delta$).  We parametrize
\eqn\lambdainter{\Delta_{jn}={j(j+1)\over k}-{n^2\over 4k}} and
define \eqn\zpsidef{\Psi=z^{|n|\over 2}e^{in\phi}F(z)\qquad
z=\rho^2 \qquad n\in \IZ.} Then $F$ satisfies the hypergeometric
equation
\eqn\hypergeoe{z(1-z)F''+[\gamma-(\alpha+\beta+1)z]F'-\alpha\beta
F=0} with \eqn\alphabeta{\eqalign{ &\alpha=\half\left(|n|+1+
\sqrt{n^2+1 +4k\Delta }\right) ={|n|\over 2} + j+1 \cr
&\beta=\half\left(|n|+1- \sqrt{n^2+1 +4k\Delta }\right)={|n|\over
2}-j\cr &\gamma=|n|+1}.}
The boundary conditions on $F$ are as
follows.  Near $z=0$ the space is smooth and from the form of
$\Psi$ \zpsidef\ we conclude that $F$ is analytic around that
point.  Therefore, $F$ is the   hypergeometric function.  The
normalization condition on $\Psi$ is that $\int \sqrt
ge^{-2\Phi}|\Psi|^2\sim \int dz z^{|n|} F^2$ converges, and
therefore we impose that $F(z=1)$ is finite.  For $\gamma >\alpha
+\beta$ $F(z=1)={\Gamma(\gamma)\Gamma(\gamma-\alpha-\beta)
\over \Gamma(\gamma-\alpha)\Gamma(\gamma-\beta)}$.  In our case
$\gamma=\alpha+\beta= |n|+1$ and we have to consider the limit:
$F(z=1)=\lim_{\epsilon\to 0}{\Gamma(|n|+1)\Gamma(\epsilon) \over
\Gamma({|n|\over 2} -j +\epsilon)\Gamma({|n|\over 2}+ j+1)}$. It
is finite only when
\eqn\lambdainterf{j={|n|\over2}, {|n| \over
2} +1,...\qquad n\in \IZ}
 or equivalently $j=0,\half,1,...$ and $-j\le {n\over 2} \le j$. In
this case $F$ is a  polynomial in $z$ of degree $j-{|n|\over 2}$.
See appendix D for further discussion of the eigenfunctions.

\subsec{Orbifolds of the parafermion theory and T-duality}

The parafermion theory has a $\IZ_k$ global symmetry
under which the fields $\psi_{j,n}$ generating the
representation $(j,n)$ transform as
\eqn\globsymm{ g: \psi_{j,n} \to \omega^n \psi_{j,n} \qquad\qquad
\omega = e^{i2 \pi/k}}
(Another $\IZ_2$ symmetry is $\psi_{j,n} \to \psi_{j,-n} $.)
Therefore, we can orbifold by any discrete subgroup of $\IZ_k$.
These are the $\IZ_l$ subgroups generated by $g^{k/l}$ for $l$
that divides $k$. Taking a symmetric orbifold by $\IZ_l$ of
\paraparti\ leads to the partition function
\eqn\paraparto{Z=\half \sum_{{j n \overline n\atop {n-\overline
n=0\mod 2l \atop n+\overline n=0\mod {2l'}}}} \chi_{jn}\overline
 \chi_{j\overline n},}
where $l'={k\over l}$. One derives this by imposing the
projection in the untwisted sector and then imposing modular
invariance.  Using the fact that $\chi_{j,n}=\chi_{j,-n}$ we see
that the partition function of the orbifold of the theory by
$\IZ_l$ is the same as that of the orbifold by $\IZ_{l'}$,
suggesting the models are equivalent.  Note, in  particular, that
the level $k$  parafermion theory would be equivalent to its
$\IZ_k$ orbifold. We will now argue that the models are
equivalent.

The equivalence of the parafermion theory to its $\IZ_k$ orbifold
can in fact  be seen at the sigma model level. Consider the the
nonlinear sigma model description of the theory based on the
metric and dilaton \parametric. A simple way to see this is to
perform T-duality on the $U(1)$ isometry of the metric.\foot{
Strictly speaking this is not correct since the system does not
really have a $U(1)$ symmetry, for reasons we explained above.
A better way to derive this equivalence is by performing the
usual steps that implement a T-duality (as in \buscher ) in the
gauged WZW model \gaugedwzw , before integrating over $A$. This
also proves it to all string loops. The net result is the same as
performing the naive T-duality.}

The resulting sigma model is based on the metric and string
coupling
\eqn\metricstring{\eqalign{
&(ds')^2={k\over 1-\rho'^2}(d\rho'^2+\rho'^2d\phi'^2)\cr
&g'_s={g_s(0)\over \sqrt k }(1-\rho'^2)^{-\half }\cr
&\rho'=(1-\rho^2)^\half\cr &\phi' \sim \phi' + {2\pi \over k}.}}
In other words, the original model with string coupling $g_s(0)$
is T-dual to its orbifold with string coupling $g_s(0)\over
\sqrt{k}$.  Note that the transformation on $\rho$ exchanges the
boundary of the disk and its center. A similar T-duality
transformation establishes the more general relation mentioned
above between a $\IZ_l$ orbifold of the theory and its $\IZ_{l'}$
orbifold.

This T-duality also sheds light on the lack of global $U(1)$
symmetry of the theory.  The metric and dilaton of the theory
\parametric\ are invariant under $U(1)$. How is it that the theory
is not invariant under $U(1)$?  The point is that close to $\rho
=1$ the sigma model description is singular because the metric
and the string coupling diverge there.  The T-duality
transformation allows us to explore this strongly coupled
region.  In the T-dual variables \metricstring\ this region is
mapped to  $\rho'=0$, where the only singularity is an order $k$
orbifold singularity of the metric. The putative $U(1)$ symmetry
associated with momentum around the disk is mapped to winding
symmetry around the origin. The orbifold singularity makes it
clear that winding is conserved modulo $k$.  We conclude that the
T-dual variables exhibit the breaking of the original $U(1)$
symmetry to $\IZ_k$ near the boundary of the disk where the
original variables are strongly coupled.

\newsec{D-branes in the parafermion theory}

\subsec{$U(1)_k$ theory}

This theory is a useful warm up example. The D-branes in this
theory are well-known (see for example, \fs).  We find here two
kinds of boundary states A-states and B-states which are
annihilated by $J(\sigma)\pm \bar J(\sigma)$ ($\sigma$ is a
parameter around the boundary).  The terminology of A-branes and
B-branes is as in $N=2$ superconformal field theories \ooy\ (see
below).  The Ishibashi A states which preserve the whole chiral
algebra are $|A, r,r\rangle\rangle$ with $r=-k+1,-k+2,...,k$. The
two integers denote the momentum of the left moving and the right
moving $U(1)$ algebras. For clarity, we give the explicit
expressions for the A,B-type Ishibashi states. The A-Ishibashi
states are
\eqn\aishi{ \vert A r,r\rangle\rangle = \exp[+\sum_{n=1}^\infty
{1\over n} \alpha_{-n} \tilde \alpha_{-n} \bigr]\sum_{\l\in \IZ}
\vert {r+2k\l\over \sqrt{2k}}, {r+2k\l \over \sqrt{2k}}\rangle}
while the B-Ishibashi states are
\eqn\bishi{ \vert B r,-r\rangle\rangle = \exp[-\sum_{n=1}^\infty
{1\over n} \alpha_{-n} \tilde \alpha_{-n} \bigr]\sum_{\l\in \IZ}
\vert {r+2k\l\over \sqrt{2k}}, - {r+2k\l \over \sqrt{2k}}\rangle }
The condition that the momentum states in the right hand side of
\bishi\ exist in the closed string spectrum is $r/k\in \IZ$; i.e.\
$r=0,k$.  It should be stressed that $|A,0,0\rangle \not=
|B,0,0\rangle$.  

Using the A-Ishibashi states we can form $2k$ Cardy states
 \eqn\uocardy{|A, \hat n\rangle_C={1 \over (2k)^{1\over 4}}
 \sum_{n'=0}^{2k-1}
 e^{- i\pi \hat n n'\over k} | A,n',n'\rangle\rangle.}
They are interpreted geometrically as $D0$ branes at $2 k$
special points on the circle.

Linear combinations of the two B states \bishi\ are the two B
Cardy states
 \eqn\uocardyb{|B,\eta=\pm 1\rangle_C=\bigl({k\over 2}\bigr)^{1/4}
 \left[ |B,0,0\rangle\rangle
 +\eta |B,k,-k\rangle\rangle\right]}
which we interpret as D1-branes with special values of the Wilson
line parametrized by $\eta$.

The B-branes can also be found by considering the T-dual theory
which is its $\IZ_k$ orbifold.  The A-branes in this theory are
obtained from \uocardy\ by summing over the images.  We find two
states corresponding to D0-branes at the two special points. After
T-duality to the original model they are mapped to two B-branes
which can be interpreted as D1-branes with two special values of
the Wilson lines.

This theory also has other D-branes which correspond to D0-branes
at arbitrary positions around the circle and D1-branes with
arbitrary values of the Wilson line.  These more generic boundary
states are perfectly consistent but the algebraic considerations
based on the $U(1)_k$ chiral algebra do not reveal them because
they are not invariant under such a large chiral algebra but only
under a smaller subalgebra.  Specifically, they are invariant
under a chiral algebra which is generated by the current
$J(\sigma)$ and its derivatives but without the exponential of
the compact boson.  A signal of these more generic D-branes is
found by examining the spectrum of open strings on the A and B
branes mentioned above.  They all have a massless ($\Delta=1$)
open string propagating on them.  It is easy to check that this
open string leads to a modulus and therefore these D-branes
belong to a moduli space of such D-branes.

The special case of $k=1$ is particularly interesting.  Here
$U(1)_1=SU(2)_1$.  In addition to the two circles of D-branes we
discussed above (D0 at an arbitrary position and D1 with an
arbitrary Wilson line) there are more D-branes.  Here all the
D-branes are on an $S^3$.  A signal of this larger moduli space
is again found by noticing the three massless moduli on the
D-branes which are in an $SU(2)$ triplet.

\subsec{A-branes in the parafermion theory}

As in the previous example, the A-branes are invariant under the
full parafermion symmetry and are found by following the standard
procedure. The parafermion Ishibashi states  $ |A,j,n
\rangle\rangle$ are constructed from the same representation
$(j,n)\in PF(k)$ on the left and right. These Ishibashi states
are in one to one correspondence with the primaries of the chiral
algebra and they are related to the Cardy states
 \eqn\cardyPF{ |A ,\hat j, \hat n  \rangle_C = \sum_{(j,n)\in PF(k)}
 {S^{PF \  jn}_{\hat j \hat n}
\over \sqrt{ S^{PF \ jn}_{00} }} |A, j,n\rangle\rangle }
 We can
calculate the open string spectrum in the standard way using the
parafermion characters $\chi_{jn}$, their modular transformation
properties and the Verlinde formula.   We find
 \eqn\cardyopen{
{}_C\langle A, \hat j, \hat n |{q'}_c^{L_0 + \bar L_0-{c\over 12}}
|A, \hat j', \hat n' \rangle_C=\sum_{(j,n)\in PF(k)}
 N_{\hat j, -\hat n, \hat j' \hat n'}^{ j n}
 \chi(q_o )_{j n}}
where $q_o = \exp[-2\pi i/\tau]$ is the open string modular
parameter and $ q_c \equiv {q'}_c^2 \equiv
  e^{ 2 \pi i \tau } $ is the closed string
parameter.

These D-branes can be interpreted as follows.  The $k$ states with
$\hat j=0$ are D0-branes at one of the $k$ special points around
the disk. The $k [{k-1\over 2}]$ states with $\hat j={1\over 2},
1, ..., \half [{k-1\over 2}]$ are unoriented D1-branes stretched
between two of the special points around the disk separated by
$2\hat j$ segments. Finally, for $k$ even we also have $k\over 2$
branes with $\hat j={k\over 4}$ stretched between two antipodal
special points (see figure 1).

\ifig\figa{ These are various A-branes in the $k=6$ case. We have
a disk with $k$ special points along the boundary. We can see a
D0-brane at one of the special points and some  D1-branes
stretched between different special points.}
{\epsfxsize1.5in\epsfbox{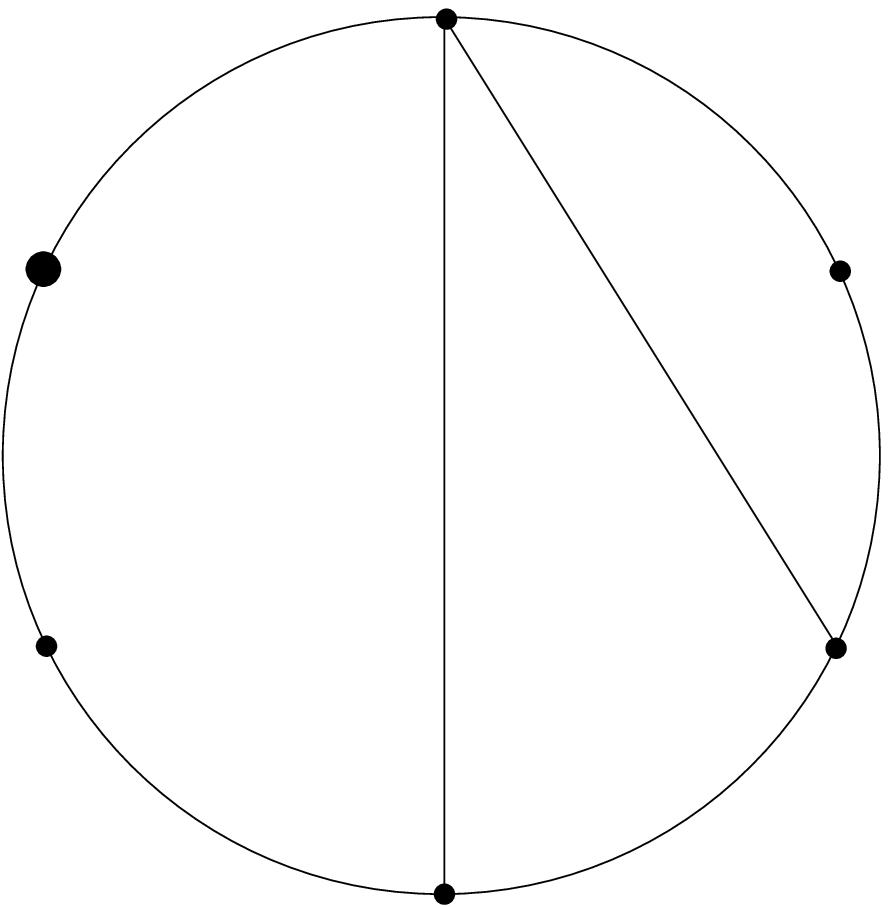}}

This interpretation is supported by the following observations:
\item{1.} We explicitly compute the shape of the branes in
appendix D by scattering closed strings.

\item{2.}  The $\IZ_k$ global symmetry
multiplies the Ishibashi states by a phase and shifts the $\hat n$
label of the Cardy states, $\hat n \to \hat n +2 $. Since this
symmetry rotates the disk, it moves a D1-brane connecting two
points separated by $2\hat j$ segments to another D1-brane
separating $2\hat j$ segments.

\item{3.}  For large $\hat j$ and $k$ the dynamics of D1-branes
can by analyzed by their DBI action.  The effective metric
for a D1-brane is
\eqn\flatmetric{{(ds)^2\over g_s^2}=k (d\rho^2+\rho^2d\phi^2).}
Since it is flat, the D1-branes are straight lines.

\item{4.}  The masses of the D-branes are found from their overlap
with $|A,j=0,n=0\rangle\rangle $, since the graviton vertex
operator in the noncompact spacetime is proportional to the unit
operator in the internal theory. Thus the masses are proportional
to
 \eqn\dbranemass{ M_{\hat j} := {1\over g_s(0)
\sqrt{k(k+2)}}\sin {\pi (2\hat j+1)\over k+2}.}
 This expression is consistent with the geometric interpretation
suggested above. For large $\hat j$ and $k$ the mass of the
D-brane is proportional to $ \sin {2\pi\hat j\over k}$, which is
the same as the length of the D-brane in the flat metric on the
disk \flatmetric. (The mass is a regularized dimension of a 
Hilbert space, in accord with the general result \tensiondimension.)

\item{5.} We can also see that if we take a conjugacy class in
$SU(2)$ that is invariant under the $U(1)$ symmetry that we are
gauging, then we can define a brane in the coset theory. This
procedure gives straight lines on the disk. To see this more
precisely we think of $S^3$ as $|z_1|^2 + |z_2|^2 =1$ the $U(1)$
we gauge is the phase of $z_2$. The disk is parametrized by
$z_1$. A conjugacy class is given by intersecting the $S^3$ by a
hyperplane in $R^4$. We get the straight lines if we intersect by
hyperplanes of the form $a z_1 = b $ where $a,b$ are constants.
These conjugacy classes are invariant under phase
rotations of $z_2$ and they project to straight lines on the disk.

\subsec{Geodesics and open strings}

As a nice check on the open string spectrum between branes given
by \cardyopen, and as supporting evidence for the above
geometrical interpretation of the states $\vert A,\hat j=0, \hat
n\rangle_C$ we will compute the action for a fundamental string
in the metric \parametric\ stretching from $(\rho=1, \phi)$ to
$(\rho=1,\phi+ {\pi n\over k})$, $n\in 2 \IZ$. Accordingly, we
compute the length of the geodesic stretching between two points
on the boundary of the disk separated by an angle $\Delta \phi$.
Such a geodesic emerges orthogonally from the boundary, reaches a
minimal value of  $z=\rho^2$, and returns orthogonally to the
boundary.

The geodesic equation in the metric \parametric\  implies
\eqn\equation{
 { \dot \rho^2 \over \rho^2} = { \rho^2 - \rho_m^2 \over
 \rho^2_m (1 -\rho^2 ) }}
where $\dot\rho = d\rho/d\phi$ and $\rho_m$ is an integration
constant which we took such that it is the minimum value of
$\rho$ along the geodesic. {}From this we find that \eqn\hafangle{
\Delta \phi/2 = { \sqrt{z_m}  \over 2} \int_{z_m}^1 dz { \sqrt{1
- z} \over z \sqrt{ z - z_m} } ={ \pi \over 2} ( 1-\sqrt{z_m} ) }

Now we can compute the length of the geodesic. We get
\eqn\length{
\ell = \sqrt{k} \sqrt{1-z_m} \int_{z_m}^1 dz { 1 \over
\sqrt{(z-z_m)(1-z)}} = \pi \sqrt{1-z_m}
}

We now use the relation between  the length and $L_0$ to find (We
use units where $\alpha' =1$)
 \eqn\relation{ L_0 = (\ell /(2 \pi) )^2
= {k \over 4} \left( 2 ({\Delta \phi \over \pi }) - ({\Delta \phi
\over \pi })^2 \right) } For an angle \eqn\deltaangle{\Delta \phi
 = \pi { n \over k} }
we get $L_0 = { n (2k-n) \over 4 k }$ in perfect accord with the
conformal dimension of the lightest open string, whose dimension
is given by the dimension of the corresponding parafermion
representation $\Delta(j=0,n)$ in the open string channel. The
conformal field theory expression is exact, and yet is a finite
power series in $1/k$. Therefore, the semiclassical analysis
should give the exact answer, as indeed it does.

\subsec{B-branes}

In this subsection we will exhibit new branes in the parafermion
theory.  We will find them using the same approach we took to
find the B-branes in the $U(1)_k$ theory above.  We will use the
fact that the theory is T-dual to its $\IZ_k$ orbifold.  A-branes
in the orbifold theory are found from the A-branes in its covering
theory using standard orbifold techniques.  Then the T-duality
maps these branes to B-branes in the original theory.

\centerline{\it Qualitative considerations}

We look for the spectrum of A-branes in the $\IZ_k$ orbifold of
the parafermion theory with string coupling $g_s(0)$. We start
with the branes described above and we add the images under
$\IZ_k$ so that the configuration is $\IZ_k$ invariant.   We find
a single state for each value of $\hat j=0,{1\over 2}, 1, ...,
\half[{k-1\over 2}]$ with mass $M_{\hat j}$ as in \dbranemass\
(we do not need to multiply the masses by $k$ for the $k$ images
because we view the target space as the disk modded out by
$\IZ_k$ and do not compare to the mass of a D-brane in the theory
on the covering space). When $k$ is even there are also two
D1-branes of mass $\half M_{\hat j={k\over 4}}$ which are
stretched from the special point to the center of the disk mod
$\IZ_k$. In terms of the theory on the covering space they are
obtained by taking $k\over 2$ images (hence the factor of $\half$
in their mass). There are two different such D-branes because
they are like fractional D-branes at fixed points of orbifolds.

This $\IZ_k$ orbifold theory with string coupling $g_s(0)/
\sqrt{k}$ is T-dual to the parafermion theory with string
coupling $g_s(0)$. Therefore, each of these D-branes should have
a counterpart in the parafermion theory.  The transformation
$\rho'=(1-\rho^2)^\half $ \metricstring\ shows that these
D-branes which are near the boundary of the disk in the orbifold
are near the center in the parafermion theory.  Therefore, these
are new D-branes not included in the list of $k(k+1)/2$ D-branes
mentioned above.

The A-brane  with $\hat j=0$ in the orbifold theory is unstable.
The open string stretched between it and its adjacent image
includes a tachyon\foot{Since we are considering a bosonic system
there are many other tachyons. In particular, the open strings
which start and end on the same D-brane always include tachyons
corresponding to functions on the space. This is the standard
instability of the bosonic string and does not signal the
instability of this brane configuration.  We will ignore it in
this discussion.}. We will see that in more detail below when we
analyze the spectrum of open strings in detail. The instability
corresponds to motion of the D0-A-brane to the center of the
orbifold.  This brane in the orbifold theory maps to a D0 B-brane
at the center of the disk. The instability corresponds to the
possibility of moving the brane off the center. This lowers the
energy since the string coupling has a minimum at the center of
the disk. Since all other B-branes are also located at the center
we expect that they should also be unstable under displacements
off the center of the disk. In terms of the A-branes in the
orbifold theory this instability is easy to understand. The
A-D1-branes with $\hat j={1\over 2}, 1, ..., \half [{k-1\over
2}]$ form, in the covering space, closed loops. They can clearly
decay to smaller closed loops which are closer to the
origin.\foot{ Some aspects of brane decay appeared while this
paper was being written \schrecent . See \rogen\ for similar
discussions in Virasoro minimal models.}

The two states with $\hat j={k\over 4}$ which exist for $k$ even
correspond in the covering space to $k\over 2$ D1-branes
stretched between antipodal points on the disk.  They cannot
decay by shrinking and therefore they are stable.

Remembering the value of $g_s(0)$ we conclude that the
parafermion theory also has the following D-brane states near the
center of the disk. There is a D0-brane with $\hat j=0$ and mass
$\sqrt k M_{\hat j=0}$.  There are D2-branes around the center for
$\hat j={1\over 2}, 1, ..., \half[{k-1\over 2}]$ with masses
$\sqrt k M_{\hat j}$.  The fact that these are D2-branes can be
seen either by following the T-duality transformation or by
thinking, for large $k$, about the way the large number of images
go through the disk mod $\IZ_k$ (see figure 2).   All these
D-branes are unstable. Finally, there are two more stable
D2-branes with masses $\half \sqrt k M_{k\over 4}$.  As is clear
from the configuration of the D1-branes in the orbifold theory,
these two D2-branes cover the whole disk.

\ifig\figb{In figure (a) we see a B-branes at the center of the
disk. It can be viewed as arising from the A-branes of the $\IZ_k$
quotient of the parafermion theory by T-duality. The covering
space of this theory is depicted in figure (b) along with a
$\IZ_k$ invariant configuration of A-branes of this theory. The
T-dual of this configuration is the D2-brane in figure (a). The
distances $\rho_m$ and $\rho'_m$ are related through
$\rho_m^2+\rho_m'^2=1$.} {\epsfxsize3.5in\epsfbox{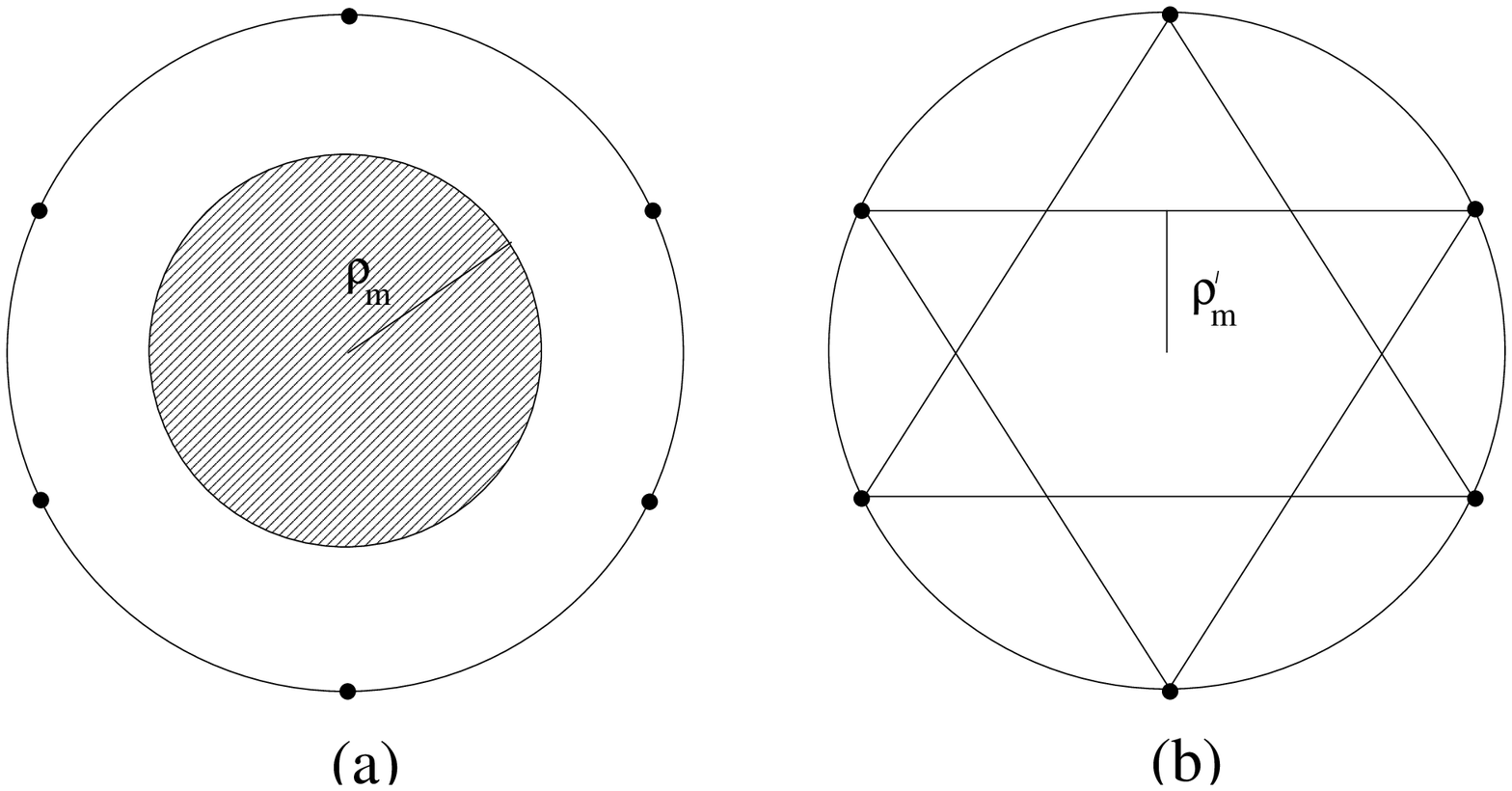}}

\centerline{\it Boundary state formulae for the B-branes}

Let us now use the above qualitative picture to give a concrete
formula for the new D2 branes centered on the disk. We must first
apply the orbifold projection to \cardyPF\ and then implement a
T-duality. The orbifold projection keeps only the $n=0$ Ishibashi
states (the $n=k$ states can be expressed as $n=0$ using the
$(j,n)\sim ({k\over 2} -j,n+k)$ identification).

Now let us implement the $T$-duality. Roughly speaking, we simply
want to change the momentum of the rightmovers relative to that
of the leftmovers. More precisely, we proceed as follows. 

The coset construction implies an equality of CFT state-spaces
\eqn\stspi{ \CH^{SU(2)}_j = \oplus_{r=0}^{2k-1} \CH^{PF}_{(j,r)}
\otimes \CH^{U(1)_k}_r }
and there is accordingly a decomposition of A-type Ishibashi
states: $\vert j \rangle\rangle = \sum_n \vert
j,r,r\rangle\rangle \vert r,r\rangle\rangle$. We now note that
the operator $e^{i \pi J^1_0}$ is block-diagonal in the
decomposition \stspi\ since
 \eqn\flip{e^{i \pi J^1_0} J^3_0 e^{-i \pi J^1_0} = - J^3_0}
Moreover, this operator may be defined to act on states in the
$U(1)_k$ theory by $\alpha_n \to - \alpha_n$ and $\vert p \rangle
\to + \vert - p \rangle$ (the choice of phase on the vacuum
is a convention). This defines, implicitly, an action
\eqn\implict{
e^{i \pi J^1_0}: \CH^{PF}_{(j,r)} \to \CH^{PF}_{(j,-r)}
}
We now define the B-type parafermion Ishibashi states by
\eqn\btypepf{ (1\otimes e^{i \pi \tilde J^1_0}) \vert j
\rangle\rangle^{SU(2)} := \sum_{r=0}^{2k-1} \vert B
j,r,-r\rangle\rangle^{PF}\otimes \vert
B,r,-r\rangle\rangle^{U(1)_k} }
It is an Ishibashi state that preserves the chiral algebra up to
a T-duality automorphism, the automorphism that changes $n\to -n$.

We may now write the boundary Cardy state for the B-branes:
\eqn\dtwostates{
|B, \hat j  \rangle_C = \sqrt{k}\sum_{ j \in Z}  { S^{PF~
 j 0}_{\hat j 0} \over \sqrt{ S^{PF~
 j 0}_{\hat 0 0} } }|B, j,0\rangle\rangle = ( 2 k )^{1/4}\sum_{ j
 \in Z}{ S_{\hat j}^{~ j}
 \over \sqrt{S_{0}^{~j} } } |B, j,0\rangle\rangle .}

There is a very interesting subtlety which arises when $k$ is
even. Below we will analyze the open string spectrum on the
D-branes. We will find that on $|B,\hat j = k/4\rangle_C$ as
defined in \dtwostates, the identity operator appears twice. The
reason for this is that this boundary state is ``reducible,''
that is, it can be written as a sum of two more fundamental
D-branes. Indeed, for even $k$ we can form another Ishibashi state:
 \eqn\newish{|B, { k\over 4},{k\over 2},-{k\over 2} \rangle\rangle
 }
where we have indicated the left and right moving $n$ quantum
numbers for greater clarity.  This B-state is in  the closed
string sector because after using the spectral flow identification
on the left part it is clear that its quantum numbers are such
that $n = \bar n$. Accordingly, we construct new irreducible D2
B-branes as
 \eqn\newirred{ | B ,{k\over 4} ,\eta \rangle_C = { 1\over 2}
(2k)^{1\over 4} \sum_{j \in Z} { S_{\hat j}^{~  j} \over
\sqrt{S_{0}^{~j} }} |B, j , 0\rangle\rangle + \eta { 1 \over 2
[k(k+2)]^{1/4}} | B {k\over 4},{k\over 2},-{k\over 2}
 \rangle\rangle,}
where $\eta=\pm 1$ and we fixed the coefficient of the last term
by demanding that we get a sensible open string spectrum.

We mentioned above that the global $\IZ_k$ symmetry multiplies
the Ishibashi state labeled by $n$ by a phase $e^{{2\pi in\over
k}}$, and therefore the A Cardy states are shifted $|A,\hat j,
\hat n\rangle _C \to |A,\hat j, \hat n +2\rangle _C$.  The B Cardy
states with $\hat j < {k\over 4}$ are clearly invariant.  But
since $|B,{k\over 4},{k\over 2},-{k\over 2}\rangle\rangle \to -
|B,{k\over 4},{k\over 2},-{k\over 2}\rangle\rangle $, the two
states with $j={k\over 4}$ transform to one another $|B,{k\over
4},\eta \rangle \to |B,{k\over 4},-\eta \rangle$.

\vskip 2cm
 \centerline{\it Properties of the B boundary states. }
\bigskip

We must now check that the states \dtwostates\ and
\newirred\ satisfy the Cardy condition and therefore
have good gluing properties. In the process we will compute the
open string spectra on these D-branes and confirm the geometrical
picture suggested above.

The overlap between two states of type \dtwostates\ is easily
computed  because the overlap between two B Ishibashi states is
the same as between two A states so we get
\eqn\dtwodtwo{\eqalign{
{}_C\langle  B\hat j & |q_c^{L_0-{c\over 24}}|B \hat j' \rangle_C
= ( 2 k )^{1/2} \sum_{ j \in \IZ} { S_{\hat j}^{~ j} S_{\hat
j'}^{~ j } \over S_{0}^{~j} } \chi_{j,0}(q_c)=\cr &
\sum_{(j'',n'')\in PF(k) } \sum_{j \in \IZ} { S_{\hat j}^{~ j}
S_{\hat j'}^{~ j } S_{j}^{~j''} \over S_{0}^{~j} }
\chi_{j'',n''}(q_o)= \sum_{j''=0}^{k/2}\sum_{n''=0}^{2k-1}
N_{\hat j, \hat j' }^{ j''} \chi_{j'',n''}(q_o)\cr} }
 (The fact that the coefficients are integers is a check on the
normalization in \dtwostates .) In deriving this formula we used
the fact that
 \eqn\sumint{
 \sum_{j \in \IZ}
{ S_{\hat j}^{~ j} S_{\hat j'}^{~ j } S_{j}^{~j''} \over S_{0}^{~j} } =
 \sum_{{\rm all~}j}
{ S_{\hat j}^{~ j} S_{\hat j'}^{~ j } S_{j}^{~j''} \over S_{0}^{j} }
{ ( 1 +
(-1)^{2j}) \over 2} = {1 \over 2}( N_{\hat j, \hat j' }^{ j''} +
N_{\hat j, \hat j' }^{ {k\over 2} -j''} )
 }
where we used $S_{j}^{~{k\over 2}-j'} = S_{j}^{~j'} (-1)^{2 j} $.
In the final step in \dtwodtwo\ we also used that the sum over
$j'',n''$ includes all parafermionic states twice.

We must now consider the overlap between the new B states and the
previously considered $\vert A \hat j\hat n\rangle_C$ states
corresponding to D0 and D1 branes. This is an involved
computation whose details are given in appendix B. The result is
\eqn\dtwodone{
{}_C\langle B ,\hat j' |q_c^{L_0-{c\over 24}}|A, \hat j \hat n
\rangle_C = \sum_{j''}
N_{\hat j', \hat j}^{j''} \tilde \chi_{j''}(q_o)
}
where   $\tilde \chi_{j}$ is a certain modular
function defined in Appendix B.

As we have mentioned, when $k$ is even there are two B branes with
$\hat j = k/4$. Their overlaps are easily computed to be
\eqn\newnewopen{ {}_C\langle B ,{k\over 4} \eta |
q_c^{L_0-{c\over 24}}| B, {k\over 4} \eta ' \rangle_C = {1 \over
4 } \sum_{j n\atop j \in \IZ} \chi_{jn} (1 + \eta \eta'
(-1)^{j-{n\over2}} ) = \sum_{ (j,n)\in PF(k)\atop {j \in \IZ \atop
j-{n \over 2} \in  2 \IZ + { 1 -\eta \eta'
 \over 2}} }\chi_{jn} }
where in the first sum we sum over all $j,n$ ($j\in \IZ$) and in
the second we restrict the sum over distinct states (i.e.\ we do
not sum over spectral flow images, as implied by $(j,n)\in
PF(k)$). Similarly we can compute the open string spectrum
involving the other B branes and we find, for $\hat j \not =
{k\over 4}$ \eqn\openstrinnew{ \langle B ,\hat j
|q_c^{L_0-{c\over 24}}|
  B ,{k\over 4} \eta  \rangle_C = { 1 \over 2 }
 \sum_{jn} N^{j}_{\hat j ,{k\over 4}} \chi_{jn} }
which is a sensible spectrum. Each state appears twice in the
sum, and therefore all states appear with integer coefficients
(actually the coefficients are one).

The overlaps of the irreducible B branes with $\hat j = k/4$ with
the A  branes can be computed using techniques described in
appendix B, but we have not carried out the details.

The above overlaps show that the Cardy conditions are satisfied,
as expected from the orbifold construction. The fact that we
obtained the branes from an orbifold construction ensures that
they obey all open strings consistency conditions, so that they
are good boundary  states.

Finally, we consider the spacetime masses of these branes. Notice
that the mass of the B states with Cardy spin $\hat j$ is the
same as $\sqrt{k}$ times the mass of the A state with Cardy spin
$\hat j$, again in perfect accord with the orbifold construction.

In appendix C we work out explicitly some examples of this
construction for low values of $k$.

One can also consider branes in various orbifolds of the
parafermion theories. It is interesting to note that in the
$\IZ_l$ orbifold of the parafermion theory the first $l-1$
B-branes become stable. For example consider the B-brane with
$\hat j=0$.  It is a D0-brane at the center of the disk. In the
original parafermion theory this brane is unstable under small
displacements from the center, since the center is where the
string coupling is smallest. If we orbifold by $\IZ_2$ this brane
becomes stable since the projection removes the mode that would
lead to the instability.

\newsec{Other descriptions of the B-branes}

\subsec{Bound states of D0-branes}

In the case of D-branes in the WZW model, for $k\gg 1$ we can
interpret the spherical D2-branes for  $ \hat j\ll k$ as states in
the theory of $N= 2 \hat j +1$ D0-branes.  These D0-branes are
described in terms of $N\times N$ matrices with non-commutative
expectation values \myers.

A similar description is possible in the parafermion theory. In
fact this was done in \disksref\ in a slightly different context.
The potential for $N$ D0-branes at the center of the disk has the
form (up to numerical constants)
\eqn\act{
  V  \sim  \int Tr\left[ - {1\over k}
(X_1^2 + X_2^2)  - [X_1,X_2]^2 + \cdots \right]
 }
where $X_1$ and $X_2$ are $N\times N$ dimensional matrices.  The
first term comes from expanding the dilaton potential $\sim
\sqrt{1-\rho^2}$.  They have a minus sign because the dilaton has
a minimum at the origin. The last term is the usual commutator
term that is present also in flat space. We neglected higher
order terms. A solution to the equations of motion is
 \eqn\sol{
X_1 \sim { 1 \over \sqrt{k}}  J_1 ~~~~~~~ X_2 \sim { 1 \over
 \sqrt{k}} J_2 }
where $J_1 $ and $J_2$ are $SU(2)$ matrices in an $N$ dimensional
representation. The D2 disks we considered above correspond to
taking irreducible representations of spin $\hat j$ ($N=2\hat
j+1$). Notice that these are not stable minima as the potential
\act\ is unbounded below. We see that the description of these
states is very similar to the description of the D2 states in
$SU(2)$. In this way of viewing them it is clear that the disks
are flattened spheres and have two sides.

\subsec{Effective action of D2-B-branes}

In this subsection we will show that B branes with $\hat j \gg 1$
can be viewed as flattened disks with induced D0 charge. The disk
has two sides and can be viewed as a flattened two-sphere.

We consider the classical equations for a disk D2 brane with an
$F$ field on it. We imagine  a fixed flux of $F$ on the disk
 \eqn\quantflux{ {N \over k} = {(2 \hat j +1) \over k} =2  {1
 \over 2 \pi}\int_{D_2} F/k = { 1 \over  \pi } \int_0^{\rho_m}
 d\rho f(\rho)~,~~~~~~~ { 2 \pi F_{\rho \varphi} \over k} = f(\rho)
  }
(we use units with $\alpha'=1$).  The factor of $2$ in the third
expression arises since our disk is made of two overlapping
branes -- a squashed $S^2$. Holding this fixed we want to minimize
the DBI action. So we minimize
 \eqn\bi{ S - \lambda (\int F -N) =
{ k \over {2 \pi} } \int d \varphi \rho \sqrt{ 1 -\rho^2}  \sqrt{
 \rho^2/(1-\rho^2)^4 + f^2 }   - k  \lambda ( \int f - {N \over k})
}
where $\lambda$ is a constant Lagrange multiplier.
The equation for $f$ that we get from this minimization procedure is
\eqn\equationf{
f = { \rho \over (1-\rho^2) } { \sqrt{ 1 - \rho_m^2 } \over \sqrt{
\rho_m^2 - \rho^2 }}
 }
where $\rho_m$ is the maximum value of $\rho$.  The Lagrange
multiplier is a function of $\rho_m$ which is of no importance.
We determine $\rho_m$ through the condition \quantflux\ and we get
\eqn\maxrho{ \int_0^{\rho_m} d\rho f(\rho) = \theta_m ~~~~~~~
\tan \theta_m = { \rho_m \over \sqrt{1-\rho_m^2} }.
 }
 So we see that $\theta_m$ is related to $\hat j$ by
\eqn\relat{
 { (2 \hat j +1) \over k } = { \theta_m \over  \pi }.
 }
We see that as $\rho_m \to 1$ then $\hat j \to {k\over 4}$ and
$\rho_m \to 1$ so that the brane is covering the whole space.
Note that from \equationf\ this brane has zero $F$, but the
integral of $F$ is still nonzero! (There is an order of limits
problem at the boundary).

We can  calculate the open string metric and open string coupling
using the transformation formulas of \sw\ (taking $B \to B+F$ and
setting $B=0$) we get
\eqn\openstr{\eqalign{
&ds^2_{open} = k { d\rho^2 + \rho^2 d\varphi^2  \over \rho_m^2 -
\rho^2 } = k {d {\rho'}^2 +  {\rho'}^2 d\varphi^2  \over 1 -
{\rho'}^2}\cr &e^{D} \equiv
G_s(\rho) =G_s(0)(1-{\rho'}^2)^{-{1\over 2}}\cr
 &\rho' = {\rho \over \rho_m} .} }
We see that the open string metric and coupling on the B-brane
are the same as the original closed string parameters.  In
particular, they are independent of $\rho_m$, which parametrizes
the size of the brane.

We conclude that the B-brane ranges from the origin $\rho=0$ to
$\rho=\rho_m=\sin \theta_m=\sin {(2\hat j+1)\pi \over k}$. This
conclusion can also be reached by considering the T-dual picture
we originally used to derive the existence of these branes.  In
this picture the target space is a wedge and these branes are
D1-branes (except for $\hat j=0$). In the covering space these
D-branes are made out of $k$ D1-branes (except for $\hat
j={k\over 4}$ when there are only $k\over 2$ D1-branes) stretched
between points at an angle of $4\hat j \pi \over k$ (see figure
2). These D1-branes are in an annulus ranging from
$\rho=\cos({2\hat j \pi \over k})$ to $\rho=1$.  Using the change
of variables \metricstring\ $\rho \rightarrow \sqrt{1-\rho^2}$ in
the T-duality, we learn that these D-branes are D2-branes ranging
from $\rho=0 $ to $\rho=\sin({2\hat j \pi \over k})$ in the
parafermion theory. This agrees with the expression for $\rho_m$
derived above (recall that this discussion makes sense only for
large $\hat j$ and $k$).

Using this metric and dilaton we should be able to find the
spectrum of small fluctuations on the D2-branes as follows.  We
consider a field $\Psi$ moving under the influence of the
Lagrangian
 \eqn\psilagopen{\sqrt Ge^{-D}{G^{ab}\partial_a \Psi\partial_b\Psi}}
with the open string  metric and dilaton given in \openstr. The
small fluctuations of $\Psi$ are controlled by the eigenvalue
equation
 \eqn\psieigeno{[-\nabla^2 + \nabla D\nabla -\Delta]\Psi=0.}
 The relative factor of $4$ between $\Delta$ here
and in \psieigen\ is common in comparisons of open strings and
closed strings.  In terms of $z={\rho'}^2$ we find that
\eqn\solupsio{\eqalign{ &\Psi=z^{|n|\over 4}e^{i{n\over
2}\phi}F(\alpha,\beta; \gamma;z)\qquad n\in 2 \IZ\cr
&\alpha=\half\left({|n|+1\over 2}+ \sqrt{{n^2+1\over
4}+k\Delta}\right) ={|n|\over 4} +{j\over 2} + {1 \over 2} \cr
&\beta=\half\left({|n|+1\over 2} - \sqrt{{n^2+1\over
4}+k\Delta}\right) = {|n|\over 4} - {j\over 2}\cr
&\gamma={|n|\over 2} +1} }
 solves \psieigeno\ where $F$ is a hypergeometric function and
we parametrized
 \eqn\lambdaintero{\Delta_{jn}={j(j+1)\over
k}-{n^2\over 4k} .}

In order to find the eigenvalues we need to impose a boundary
condition. It is not clear which is the correct boundary
condition. If we impose the boundary condition that $\Psi$
vanishes at $\rho'=1$, then we get that $j = { n \over 2} + 2 s$,
$s \in \IZ$. On the other hand if we impose that the radial
derivative vanishes at $\rho' =1$, then we get $j ={ n \over 2} +
2 s +1$. We see from \dtwodtwo\ that the B branes  with $\hat j <
k/4$ have open string states with these two sets of eigenvalues.
It is tempting to interpret these two possible boundary condition
as the ones that would result from taking wavefunctions that are
either symmetric or antisymmetric under the exchange of the two
disks that form the B branes.

In the particular case of $\hat j = k/4$ where we have a brane
wrapping the whole manifold, we have only one copy of the brane,
and therefore we have only one set of eigenfunctions, the one with
$j = n/2 + 2 \IZ$,  as we can see from \newnewopen.

As in the case of closed strings, we get  $|n/2| \leq j$.
Parafermions with $n$ outside this range are interpreted as
massive string states.

\newsec{D-branes in the superparafermion theory}

\subsec{Qualitative description}

The ${\cal N} =1$ supersymmetric version of parafermions turns
out to have also $\NN=2 $ supersymmetry. In fact they are  the
$\NN=2$ minimal models. Some aspects of these models are
summarized in appendix E.

We can think of the model as given by the coset
\eqn\supercoset{{SU(2)_k\times U(1)_2\over U(1)_{k+2}}~~~~.}
 The states in this model are parametrized by $(j,n,s)$ where $j$
is an $SU(2)_k$ spin, $n$ is a $U(1)_{k+2} $ label (i.e.\ it is an
integer $\mod 2(k+2)$), and $s$ is a $U(1)_2 $ label. The labels
should be such that  $ 2j + n + s $ is even. The labels $(j,n,s)$
and $({k \over 2 } - j, n + k+2, s+2)$ describe the same
state\foot{ We sometimes call this change in labels ``spectral
flow''. This should not be confused with the spectral flow of the
${\cal N}=2$ theory that changes NS to R states.}. We denote the
set of distinct labels by $SPF(k)$. States with $s$ even are in
the NS-NS sector and states with $s$ odd are in the RR sector. If
we consider the diagonal modular invariant of \supercoset, we do
not have NS-R sectors. If we are interested in the superstring, we
would have to include them. In this section we will restrict the
discussion to the model \supercoset\ with the diagonal modular
invariant.

The theory has a discrete symmetry $G$ (see appendix E) generated
by $g_1$ and $g_2$ under which the states transform as
 \eqn\discretesym{\eqalign{
 &g_1\Psi_{j,n,s} = e^{2\pi i\left({n\over 2k+4}-
 {s\over 4}\right)} \Psi_{j,n,s} \cr
 &g_2\Psi_{j,n,s} =  (-1)^s \Psi_{j,n,s} .}}
Of particular interest to us is the $H=\IZ_{k+2}\times\IZ_2$
subgroup generated by $g_1^2g_2$ and $g_2$.

There are $2(k+2)(k+1)$ Ishibashi states and
the same number of Cardy states labeled by $(\hat j, \hat n, \hat
s)$. These states satisfy A-type boundary conditions.  Cardy
states with $\hat s =0,2$ obey the boundary conditions\foot{We
use open string notation.  In the closed string channel there is
a factor of $i$ in the first equation and a minus sign in the
second.} $G^\pm=\bar G^\mp$, $J=-\bar J$. We call these even
A-boundary states. Their expansion contains both NS-NS and RR
Ishibashi states. The  Cardy states with $\hat s=\pm 1$ satisfy
$G^\pm=-\bar G^\mp$. We call these odd A-boundary states. Both
types of  Cardy states leave unbroken a chiral algebra isomorphic
to that of \supercoset.

As a sigma model this is an $\NN=2$ sigma model whose target space
is the disk with our familiar metric with fermions in the tangent
space.  Using a chiral redefinition these fermions can be made
free and can be bosonized to a string scale circle.  In terms of
the coset \supercoset, this circle represents the $U(1)_2$
factor.  The full symmetry group of the theory $G$ acts on the
disk times the circle. We would like to have a geometric
description of the D-branes on the disk.  To do that we have to
project the circle onto the disk. Clearly, there is a slight
ambiguity in doing that. We find it convenient to project it such
that a $\IZ_{2k+4}\subset G$ acts geometrically by rotations. Then
there are $2k+4$ special points at the boundary of the disk.  We
break them to two groups: the even points and the odd points.

The $(k+2)(k+1)$ even Cardy states  are interpreted as oriented
D1-branes stretched between the even points, and the
$(k+2)(k+1)$ odd Cardy states  are interpreted as oriented
D1-branes stretched between the odd points. Unlike the bosonic
problem, there are no D0-branes at the special points. More
precisely, the geometrical interpretation of A-type Cardy states
is that they are oriented straight lines connecting angles
$\phi_i$ to $\phi_f$ on the disk. For a state labeled by  $( \hat
j,  \hat n,  \hat s)$ we have
\eqn\astate{
\phi_i = (\hat n-2\hat j-1){\pi\over (k+2)}  \qquad\qquad
\phi_f = (\hat n+2\hat j+1){\pi\over (k+2)}
}
for $\hat s=0,-1$.  For $\hat s=+1,2$ the state is oriented from
$\phi_f$ to $\phi_i$. It is a pleasant exercise to check that
this is compatible with the state-identification. \foot{If we
choose the branch $-\pi\leq \phi\leq \pi$ for our angles then we
should choose the fundamental domain $2j-k-1\leq n\leq k-2j+1$ for
the labels on the representations.} Thus, even type A-branes
connect odd multiples of ${\pi\over (k+2)}$ around the circle,
while odd type A-branes connect even multiples of ${\pi\over
(k+2)}$.

The masses of these states are $M_{\hat j}\sim \sin{\pi(2\hat
j+1) \over k+2}$.  We note that unlike  the bosonic problem,
here the expression for the mass is exactly the length of the
D1-brane and there is no finite shift of order $1\over k$ or
$1\over \hat j$. In \HIV\ very similar  branes were studied using
the massive Landau-Ginsburg description and some topological
quantities were computed. Their results are very reminiscent of
ours, but the precise relation between them remains elusive.

We could generalize these boundary conditions and interpolate
between these states by considering $G^\pm=e^{\pm i\alpha}\bar
G^\mp$. These more general boundary conditions break $N=1$
supersymmetry but preserve all other consistency conditions. For
each of the D1-branes mentioned above there is an $S^1$ moduli
space of D-branes parametrized by $\alpha$.  This $S^1$ is the
circle of the bosonized fermions corresponding to the $U(1)_2$ in
\supercoset. Two points on this moduli space are special because
they preserve the chiral algebra of \supercoset\ and a fixed $N=1$
subalgebra. This is the familiar spectral flow interpolation
between NS and R states. The open strings between these Cardy
states would be in representations of the $\NN =2$ spectral
flowed algebra. The open string spectrum between a brane with any
$\alpha$ and itself preserves a full $\NN=2$ algebra and
therefore contains a massless state in the open string sector
associated to the $U(1)$ current in the $\NN=2$ algebra,
$J_{-1}$. This is the modulus associated to changes in  $\alpha$.
We will not discuss this further  here.

Let us now mod out this theory by the  $\IZ_2$ symmetry generated
by $g_2$ in \discretesym.   This projects out all the RR closed
string states. The twisted sector states are also  RR states with
the opposite fermion number. The partition function of the
original model is $Z_{diag}=\half \sum_{j,n,s} |\chi_{j,n,s}|^2$
and is related to the partition function  of the quotient theory
by
\eqn\quotthry{ Z' = \half \sum_{j,n,s} \chi_{j,n,s}\overline
 \chi_{j,n,-s} = Z_{diag} - (k+1) }
The  additive constant comes from  the Ramond sector states with
$L_0 = c/24$. Indeed we have $\chi_{j,2j+1,1} = \chi_{j,2j+1,-1}
+ 1$ (together with the charge conjugate equation). Since the
partition functions are distinct the theories are not equivalent.
In fact, the D1-branes we discussed above become unoriented. One
way to see that is to note that we no longer have the primary RR
states which are annihilated by $J_0-\bar J_0$ (recall that while
the boundary condition in the open string channel is $J=-\bar J$,
it is $J=\bar J$ in the closed string channel). Therefore, in the
orbifold theory, the even Cardy states with $\hat s=0$ and $\hat
s=2$ are the same and the odd Cardy states with $\hat s=\pm1$ are
also the same.

We can now further  mod out this model by $\IZ_{k+2}$; i.e.\ mod
out the original model by $H=\IZ_{k+2}\times \IZ_2$. Now the
partition function is $\half \sum \chi_{j,n,s}\chi_{j,-n,-s}$
which is equal to $Z$. This model is T-dual to the original one.
The D1-A-branes in this model are determined in a way similar to
the bosonic problem.  There is one D1-brane for each value of
$\hat j=0,{1\over 2},...,{1\over 2}[{k-1\over 2}]$ with mass
$M_{\hat j}$, and for even $k$ two more D1-branes going through
the center of the orbifold with mass $\half M_{\hat j={k\over
4}}$. Since this model is T-dual to the original model (up to a
rescaling of $g_s(0)$ by $\sqrt {2(k+2)}$), the super-parafermion
theory should also have B-branes labeled by $\hat j=0,{1\over
2},...,{1\over 2}[{k-1\over 2}]$ with masses $\sqrt{2(k+2)}
M_{\hat j}$ and for even $k$ two more B-branes covering the space
with mass $\half \sqrt{2(k+2)} M_{\hat j={k\over 4}}$.  These are
D2 branes at the center of the disk, since the B-brane with $\hat
j=0$ is small we can also refer to it as a D0-brane.  Most of
these B-branes are unoriented; i.e.\ they are not accompanied by
anti-B-branes with opposite quantum numbers. The two B-branes
with $\hat j={k\over 4} $ can be thought of as oriented because
they transform to one another by the orbifold $\IZ_2$.

Once we have learned that these D2-branes are present in the
super-parafermion theory, we can find them also in its $\IZ_2$
orbifold.  Each of the B-branes with $\hat j=0,{1\over
2},...,{1\over 2}[{k-1\over 2}]$ leads to a D2-B-brane and an
anti-D2-B-brane, while the two D2-branes with $\hat j= {k\over
4}$ lead to a single D2-brane with that value of $\hat j$.

The original superparafermion theory and its $\IZ_2$ orbifold are
T-dual to two other models (their $\IZ_{k+2}\times \IZ_2$
orbifolds) in which the D1-A-branes and the D2-B-branes are
interchanged.

These D2-B-branes satisfy the boundary conditions $J=\bar J$,
$G^\pm=\pm \bar G^\pm$ (the sign in the latter expression depends
on whether the Cardy state is even ($\hat s=0,2$) or odd ($\hat
s=\pm 1$)).  They preserve an $\NN=2$ subalgebra which differs
from that which is preserved by the D1-A-branes. Like the
A-branes, they are also part of a one parameter family of
D-branes with the boundary conditions $G^\pm=e^{\pm i\beta} \bar
G^\pm$.  We can interpret the parameters $\alpha$ and $\beta$ as
a circle (of $U(1)_2$ in \supercoset) and its dual.  The A-branes
and the B-branes are D0-branes and D1-branes with respect to this
circle.

\subsec{More quantitative description}

The $S$ matrix of the model is
 \eqn\smatrix{ S_{j n
s}^{SPF~~~j'n's'} = 2 S_j^{j'} S_n^{n'} S_s^{s'} = {1\over
\sqrt{2(k+2)}} S_{j}^{~j'} e^{ {i \pi n n'\over k+2}} e^{- i \pi s
s'\over 2} }
 where $S_j^{j'},~ S_n^{n'}$ and $ S_s^{s'}$ are the $S$ matrices
of $SU(2)_k $, $U(1)_{k+2}$ and $U(1)_2$ respectively.  The factor
of $2$ arises due to spectral flow identification.

\centerline{\it A boundary states}

The obvious Cardy states are
 \eqn\cardya{ |A, \hat j \hat n \hat s \rangle_C = \sum_{(jns)
\in SPF(k)} { S_{\hat j \hat n \hat s}^{SPF~~jns} \over
\sqrt{S_{000}^{SPF~~jns} }} |A,jns\rangle\rangle }
The Ishibashi states have the same
left and right $s,n$. This implies that in the closed string
channel we have $J = \bar J$ and therefore $G^\pm = i \bar
G^{\mp}$ or $G^\pm = -i \bar G^{\mp}$ \foot{Notice that this is
consistent with the commutation relations of the current algebra
once we take into account that these equations only hold when
acting on the boundary state i.e.  $[J,G^\pm] |B \rangle = [i
\bar G^\mp,\bar J] |B\rangle $, i.e. the order of the operators
is reversed.}.

Then the open string partition function is obtained as
\eqn\openstr{{}_C\langle A, \hat j \hat n \hat s|q_c^{L_0- {c\over
24} } |A, \hat j' \hat n' \hat s' \rangle_C =\sum_{jns \in SPF(k)}
N_{\hat j, -\hat n, -\hat s ; \hat j' \hat n' \hat s'}^{jns}
\chi_{jns}(q_o) }
 where we sum only over distinct states. The fusion rules are
  \eqn\opecoef{ N_{\hat j, -\hat n,-\hat s ; \hat j'
\hat n' \hat s'}^{jns} = N_{\hat j, \hat j'}^j \delta_{n- \hat n
+ \hat n'} \delta_{s - \hat s + \hat s'} + N_{\hat j, \hat
j'}^{{k\over 2} -j} \delta_{n + k+ 2- \hat n + \hat n'} \delta_{s
 + 2 - \hat s + \hat s'} }
where the delta functions are periodic delta functions with the
obvious periods, in other words $\delta_n$ had period $2 (k+2)$
and $\delta_s$ period $4$ (the index distinguishes the periods).
Using \opecoef\ it is convenient to write \openstr\ as
 \eqn\openpart{ {}_C \langle A,\hat j \hat n \hat s|q_c^{L_0-
{c\over 24}} |A, \hat j' \hat n' \hat s' \rangle_C =\sum_{j}
N_{\hat j \hat j' }^{j} \chi_{j, \hat n - \hat n', \hat s -\hat
s'}(q_o). }

Let us now comment on the geometrical interpretation of
\openpart. First, we must bear in mind that the bra ${}_C \langle
A,\hat j \hat n \hat s|$ represents a brane with the {\it
opposite} orientation to the brane $ |A, \hat j \hat n \hat s
\rangle_C$. Thus,  the open string spectrum on a single brane is
computed from \openpart\ with $\hj=\hj'$ and $\hn=\hn'$, but $\hs
-\hs'=2$. Using equation $(E.12)$ of appendix E we see that there
is no tachyon on the A-brane. Hence, our A-branes are  stable
branes. Switching to $\hs =\hs'$ we find many tachyons. This is
expected for overlapping branes and anti-branes.

Next let us study  the open string spectrum for strings
connecting different branes. There is an interesting interplay
between the existence of tachyons in the open string spectrum and
the geometrical intersection of the two branes. Roughly speaking,
when two even-type or odd-type A-branes intersect, or are close
together, we find instabilities. When the branes do not intersect
and are more than a string length apart the configuration is
stable. If even-type and odd-type branes intersect the open string
is in the Ramond sector and there are no tachyons.

More precisely, suppose the brane $(\hat j,\hat n, \hat s)$ is a
straight line between angles $[\phi_i, \phi_f]$ given in \astate.
Denoting the analogous angles for $(\hat j',\hat n', \hat s')$
by  $[\phi_i', \phi_f']$ the straight lines intersect iff
\eqn\intercond{
\phi_i\le \phi_i' \le \phi_f \le \phi_f' \qquad {\rm or} \qquad
\phi_i'\le \phi_i \le \phi_f' \le \phi_f.
 }
Let us consider the first case, for definiteness. Then the three
inequalities imply that $\vert 2\hat j - 2\hat j'\vert \leq \hat
n - \hat n' \leq 2 \hat j+2 \hat j' + 2$. One easily checks that
if $\hat s-\hat s'=0$ then there is at least one  tachyon in the
open string sector (and often many more). If $\hs - \hs'=2$ there
is no tachyon. The instability signaled by these tachyons is easy
to interpret geometrically.  When two branes intersect they tend
to break and form two shorter branes which do not intersect. This
breaking must be consistent with the orientation.  When one brane
ends where another brane starts the instability is toward the
merging of these two branes into a single shorter brane.\foot{The
two intersecting branes define diagonals on a quadrilateral.
Denoting by $\ell_{\phi_1,\phi_2}$ the Euclidean length of a
straight line between angles $\phi_1,\phi_2$ it follows from the
triangle inequality that we have $\ell_{\phi_i'\phi_i} +
\ell_{\phi_f\phi_f'}\leq \ell_{\phi_i'\phi_f'} +
\ell_{\phi_i\phi_f}$ as well as $\ell_{\phi_i\phi_f'} +
\ell_{\phi_i'\phi_f}\leq \ell_{\phi_i'\phi_f'} +
\ell_{\phi_i\phi_f}$. Thus there is nonzero phase space for the
decay channel in both orientations. The presence or absence of
tachyons dictates whether this decay takes place at string tree
level or nonperturbatively.}

Now let us consider the case where the branes do not intersect.
In this case either $\phi_i < \phi_f < \phi_i'<\phi_f'$ or
$\phi_i < \phi_i' < \phi_f'<\phi_f$ (up to exchange of $\phi$
with $\phi'$). In the first case there are no tachyons. In the
second case there can be  tachyons even when the branes do not
intersect. Nevertheless, one can check using equation $(E.12)$
that this happens only when the nonintersecting branes are
separated by a distance of order the string scale, in the string
metric.

The decays mentioned above suggest that one can introduce a
lattice of conserved charges. For the even A-type states we have
$\IZ^{k+2}/\IZ \cong \IZ^{k+1}$, and for the odd A-type states
another copy of the same lattice. The quotient by $\IZ$ accounts
for the fact that a ring of A-type states can shrink into the
disk. \foot{We thank M. Douglas for a useful discussion on this.}
There are interesting relations of this lattice to equivariant
K-theory and to the algebra of BPS states, but we will leave this
for another occasion.

\centerline{\it B-boundary states}

We can similarly find the $B$ branes. Let us first find the
branes for generic $\hat j$ and then discuss the branes with
$\hat j = {k\over 4}$ for even $k$.

We can build the generic branes as outlined above. We start with
branes in a PF' model with partition function $\sum \chi_{j,n,s}
\bar\chi_{j,-n,-s}$. Then we quotient these branes by $\IZ_{k+2}
\times \IZ_2$ to get branes in the original model, i.e. $PF =
PF'/(\IZ_{k+2}\times \IZ_2)$. The Ishibashi states of the  $PF'$
model  are $ |B, j, n, s ;j,-n,-s\rangle\rangle$.  The only
Ishibashi states that exist also in the $PF$ model are the ones
with $n=0,k+2$, $s=0,2$ and any $j$. Using the spectral flow
identification it is enough to consider only $n=0$ with $s=0,2$.
For even $k$ we can also have the states $ |B, {k \over 4}, { k+2
\over 2} , \pm 1 ; {k \over 4}, -{ k+2 \over 2} , \mp 1
\rangle\rangle$ . The label B reminds us that these states  obey
different boundary conditions for the supercurrents so that they
are not identical to the A Ishibashi states.

If we start with Cardy states similar to \cardya\ in the $PF'$
model and then we add all the images under $\IZ_{k+2}\times \IZ_2$
we get a state that is invariant under $\IZ_{k+2}\times \IZ_2$ and
is therefore in the orbifold theory. The resulting  state contains
Ishibashi states with only $s=0,2$ and $n =0$.
 \eqn\cardybfirst{|B ,\hat j ,\hat n , \hat s \rangle_C = \sqrt{
 2 (k +2)} \sum_{j \in \IZ , s=0,2} {{S^{SPF}}_{\hat j, \hat n,
 \hat s}^{~~~~j 0 s} \over \sqrt{ {S^{SPF}}_{000}^{~~j 0 s} } }
 |B, j,0,s;j,0,-s\rangle\rangle }
The factor of $\sqrt{2(k+2)}$ arises from normalizing the open
string channel. Alternatively, it can also be viewed as a
standard factor arising whenever we do orbifolds. Note the weak
dependence on the labels $\hat n$,$\hat s$. They simply serve to
enforce the selection rule.
Since we are summing only over $s$ even, only the
values of $\hat s =0,1$, which correspond to $G = \pm \bar G$ on
the boundary, produce distinct states. So the total number of
states is $  2 [{k\over 2}]+1  $ (not counting the factor of 2
coming from the two signs in the $G$ boundary condition). These
come from $\hat j =0,{1\over 2},....[{k\over 2}]/2$.  (We will
later see that for $k$ even the last of  them is reducible). These
branes are not oriented because there are no RR Ishibashi states.

We can now compute the open string spectrum between these branes
\eqn\openbstates{
 {~}_C\langle B ,\hat j, \hat n ,\hat s|q_c^{L_0-{c\over 24} }
|B ,\hat j',\hat n', \hat s' \rangle_C
= \sum_{(j,n,s)\in SPF(k)} (N_{\hat j\hat j'}^{~~j}
+N_{\hat j\hat j'}^{~~{k\over 2}-j} ) { (1+(-1)^{\hat s' -\hat s-s})
\over 2 }
\chi_{(j,n,s)}(q_o)
}
We see that the identity open string state appears once in the
diagonal matrix elements except for the case $\hat j = {k\over
4}$ for $k$ even when it appears twice.  This is an indication
that the B-brane with $\hat j = {k\over 4}$ is reducible.

Let us now study this case in more detail. From now on $k$ is
even. So there is an extra Ishibashi state that we did not use in
\cardybfirst\ . In order to keep notation simple we define the
state $| B, \hat s\rangle_C$ to be the state in \cardybfirst\
with $\hat j = {k\over 4}$. The new states are
 \eqn\cardybnew{ |B, {k \over 4}
 \hat s \rangle_C = { 1 \over 2}\left( |B, \hat s\rangle_C
+ \sqrt{k+2}
e^{-i {\pi  \hat s^2 \over 2} }
\sum_{s=1,-1}  e^{- i
{\pi {\hat s} s\over 2}} |B, {k \over 4} , { k+2 \over 2}, s;{k
\over 4} , -{ k+2 \over 2}, -s \rangle\rangle\right),
 }
where the coefficient of the extra state was fixed by demanding a
reasonable open string partition function. The factor involving
$\hat s^2$ is necessary in order to get some $i's$ appearing
appropriately. There are four distinct states labeled by $\hat s$.
The branes \cardybnew\ are oriented because there are RR
Ishibashi states.

Then the open string partition function between the states
\cardybnew\ is
 \eqn\openbnew{ {}_C\langle B {k \over 4}
\hat s |q_c^{L_0-{c\over 24} }| B {k \over 4} \hat s' \rangle_C =
\sum_{n,j} \chi_{j, n , \hat s' - \hat s } { 1\over 2} [ 1 +
(-1)^{( \hat {s'}^2 + \hat s' - \hat s^2 -\hat s)/2}
 (-1)^{ (2j + n - \hat s + \hat s' )/2} ]
}
The sum is over $j,n$ such that $j\in \IZ$ and $n+\hat s' -\hat s
= 0 \mod 2$.  All states have coefficient one. Notice that the
quantities in the exponent are integers due to the selection rule.
Putting $\hat s = \hat s'+2$ we see that these branes are stable.

We can similarly compute the open string spectrum between one of
these branes and the other B-branes. In this case the new
Ishibashi states in \cardybnew\ do not contribute and we just get
half the contribution in \openbstates , but with $\hat j =
{k\over 4}$. It is easy to check that for this case all states
appear twice in \openbstates , so that when we take half of
\openbstates, with $\hat j={k\over 4}$, we get a sensible open
string spectrum. This open string spectrum only depends on the
value of $\hat s$ in \cardybnew\ modulo 2, which implies that we
get the same spectrum for the brane and the antibrane at $\hat j
= {k\over 4}$.

The open string spectrum between the A-type and B-type branes is
computed  in appendix B.

\subsec{Remarks concerning the Witten index}

It is interesting to note that the $U(1)$ current in the $\NN=2$
algebra is
 \eqn\ntwo{ J = { s \over 2} - { n \over k+2} .}
The NS sector chiral primaries are given by $\phi_{l,-2 l,0}$ (and
antichiral primaries given by $\phi_{l,2l,0}$), with $l=0,{1\over
2},... {k\over 2}$. So there are $k+1$ chiral primaries. They are
also equivalently given by the labels related by spectral flow to
these. Spectral flow from the NS to the RR sector is given by the
change $(n,s) \to (n+1,s+1)$ in the labels of the field. In the
NS sector the spectrum of  $J$ for chiral primaries is bounded as
$ 0 \leq J \leq c/6 $. The chiral primaries in the R sector are
$\phi_{l,2l+1,1}$.

It is interesting for some purposes to compute the Witten index
$\Tr_R(-1)^F q^{L_0}_o$ in the open string R sector. In order to
do this computation we need to have a boundary state with $G=
\bar G$ boundary condition on one side and $G = - \bar G$
boundary condition on the other side of the strip. This will
ensure that the open string has R boundary conditions which
preserve supersymmetry. Given the Cardy state $|A,\hat j, \hat n,
\hat s \rangle_C$ we produce a state with opposite boundary
conditions by considering $|A,\hat j, \hat n + 1, \hat s +1
\rangle_C$. Notice that we need to shift $\hat n$ due to the
selection rule. We want to think about them, however, as
describing the same brane.  The Witten index was computed in
\refs{\HIV,\bdlr}, we just reproduce their calculation  and
include the index between A branes and the new B brane.

We also need to introduce the factor of $(-1)^F$ in the
computation. This factor will imply that only RR states propagate
in the closed string channel. We know that if we have the Cardy
state with label $\hat s$, then we reverse the sign of the RR
sector in the closed string channel by shifting $\hat s \to \hat
s+2$. Then we conclude that the Witten index is
 \eqn\wittenindex{ \Tr_{\CH_{\alpha\beta}}(-1)^F = {}_C\langle A
 \hat j ,
\hat n, \hat s |q_c^{L_0-{c\over 24} }| A \hat j' , \hat n', \hat
s' \rangle_C -   {}_C\langle A  \hat j , \hat n, \hat s
|q_c^{L_0-{c\over 24} }| A, \hat j' , \hat n', \hat s'+2
\rangle_C,
 }
where $\hat s -\hat s' =\pm 1 \mod(4)$  in order for the open
string Hilbert space $\CH_{\alpha\beta}$ corresponding to
boundary conditions $\alpha\beta$ to be in the Ramond
sector.\foot{Notice that there is no factor of $1/2$ in
\wittenindex\ because the overlaps of Ishibashi states have
definite fermion number, they are really $\Tr[{ (1 \pm (-1)^F )
\over 2  } q_c^{L_0}] $. So in \wittenindex\ we extract the term
with $(-1)^F$. }

Using \wittenindex\ and  \openpart\  we find that the Witten
index between two  A  branes (with $\hat s -\hat s' = \pm 1
\mod(4)$) is
 \eqn\wittenind{ \Tr(-1)^F =
(-1)^{ \hat s - \hat s' -1 \over 2} \sum_{j} N_{\hat j,\hat j'}^j
\biggl( \chi_{j,\hat n-\hat n',1}
 -\chi_{j,\hat n-\hat n', -1}\biggr)
}
 Chiral primaries can only appear in the R sector if
$ 2 j = \hat n - \hat n' - 1$ from the first term or if $ k - 2 j
= k+2 + \hat n - \hat n' -1$ in the second term. In order to make
formulas more transparent we take $\hat s =1, ~ \hat s' =0 $, and
define $\hat n = \tilde n +1$. So we think of $(\hat j,\tilde n)$
and $(\hat j',\hat n')$ as the labels of the two states. All
together we get
 \eqn\wittenindfin{ \Tr(-1)^F = N_{\hat j \hat
 j'}^{\tilde n - \hat n'} }
where, as in \bdlr\ we defined $ N_{j,j'}^{-j-1} = -N_{j,j'}^{j}$
and $ N_{j,j'}^{-{1\over 2}} = N_{j,j'}^{{k\over 2}+{1\over 2}}
=0$. Notice that only one of the two terms in \wittenind\ is
nonzero. Using \intercond\ one can show that the condition for the
existence of a chiral primary is precisely the condition for the
D1 branes to have an intersection, and that moreover \wittenind\
is simply the oriented intersection number of the two lines. This
is, once again, strongly reminiscent of the result of \HIV.

Note that the Witten index between any of the A states and the
unoriented B states is zero, as well as between any of the
unoriented B states, due to the fact that the B boundary
states have no RR sector states.

We now compute the Witten index between the oriented $B$ branes
with $\hat j = k/4$  and the $A$ branes:
\eqn\abwitten{ {}_C\langle B {k\over 4} \hat s'\vert
q_c^{L_0-c/24} \bigl(\vert A \hat j,\hat n,\hat s\rangle_C -
\vert A \hat j,\hat n,\hat s+2\rangle_C \bigr) }
This can be written as
\eqn\wittenab{
 \sin[{\pi\over 2}(2\hat j+1)]e^{i \pi (\hat s')^2/2}
\biggl( e^{{i \pi \over 2}(\hat s' + \hat n - \hat s)} f_{+}
+
 e^{{i \pi \over 2}(-\hat s' - \hat n - \hat s)} f_{-}
\biggr) }
where $f_{+},f_{-}$ are the overlaps
 \eqn\defoverlps{\eqalign{
 f_+  =\langle \langle B {k\over 4}, {k+2\over 2},1
 ;-  {k+2\over 2},- 1 |q_c^{L_0-{c\over 24}}|A {k\over 4},
 &  {k+2\over 2},1  \rangle \rangle \cr
 f_-  =\langle \langle B {k\over 4}, {k+2\over 2},- 1;-
 {k+2\over 2},+1  |q_c^{L_0-{c\over 24}}|A {k\over 4},
 &  {k+2\over 2},-1  \rangle \rangle
 \cr}}
These overlaps are related to conformal blocks of one-point
functions of simple currents on the torus, and as such have been
much studied (see, for examples,
\refs{\mooreseiberg\fss-\falceto}). However, the only key fact we
need is that the result must be $q_c$-independent and therefore
we can take the $q_c\to 0 $ limit and only keep the terms with
the RR groundstates\foot{Technically  only $f_+$ has a
contribution from the RR groundstate, and in fact $f_+= 1+ f_-$
where $f_-$ is a nontrivial function of $q_c$. There is a
cancellation in the two terms in \wittenab.}. The resulting
overlap is then immediately evaluated to be:
\eqn\abwittindx{ {}_C\langle B {k\over 4} \hat s'\vert q^H
\bigl(\vert A \hat j\hat n\hat s\rangle_C - \vert A \hat j\hat
n\hat s+2\rangle_C \bigr) = \cases{
 0 & $ \hat j \in \half + \IZ $ \cr
 (-1)^{\hat s'(\hat s'+1)/2}
(-1)^{(2\hat j + \hat n - \hat s)/2} &$ \hat j \in \IZ$\cr} }

The Witten index of boundary states can usually be interpreted in
terms of intersection theory.  The expression \abwittindx\ may
likewise be identified with  the intersection between the lines
associated with A-branes with a signed sum of points on the
boundary of the disk. For example, the intersection number
\abwittindx\ is reproduced by the intersection number of the
A-branes with the signed sum of points $(-1)^{\hat s'(\hat
s'+1)/2} (P_0 + P_1)$, where $P_0$ consists of the sum of points
at the boundary of the form $4n{\pi \over k+2}$ and $P_1$ is the
sum of points at the boundary of the form $(4n+1){\pi \over k+2}$.
This interpretation suggests the possibility that there might be a
nonperturbative instability of the special disk-filling B-brane
to a collection of oriented D0 branes on the boundary of the disk.

\newsec{D-branes in $SU(2)$ and Lens spaces}

In this section we will construct some new branes for the $SU(2)$
WZW model, these are branes which are analogous to the B-branes
we described for parafermions. In the case of parafermions, it was
useful to view the model as a $\IZ_k$ quotient, up to a T-duality.
The same idea will be useful in the $SU(2)$ case. So we first
start reviewing Lens spaces, which are interesting quotients of
$SU(2)$, and then we will proceed to discuss the branes.

\subsec{Lens spaces}

The Lens space is $S^3/\IZ_{k_1}$ where the $\IZ_{k_1}$ is a
discrete subgroup acting on the left (as opposed to the $U(1)$
discussed thus far). In terms of the coordinates in \metricsth\
it corresponds to the identification $\chi \sim \chi + {4 \pi
\over k_1}$. This is a quotient without fixed points. The Lens
space is completely nonsingular.  In terms of the $SU(2)$ WZW
model this is the orbifold $SU(2)/\IZ_{k_1}^L$ where
$\IZ_{k_1}^L$ is embedded in the left $U(1)$. In order for this
theory to be consistent, the level $k$ of the $SU(2)$ covering
theory should be of the form $k=k_1k_2$ where $k_1,k_2\in \IZ$.
This can be understood as quantization of three form $H$ in the
coset theory. In string theory this theory arises as the
transverse geometry in coincident $A_{k_1-1} $ singularities and
NS 5-branes.

The partition function for this theory was studied in \gps, where
it was written using parafermions. Using \parafdeco\ the
partition function of $SU(2)_k/\IZ_{k_1}^L$ is
 \eqn\part{\eqalign{
Z =& \sum_j \sum_{ n +n'=0 mod 2 k_1 \atop n-n' =0 mod 2 k_2}
 \sum_{\bar n = -(k-1)/2}^{{k\over 2}} \chi_{j n}^{PF}(q)
\chi^{U(1)}_{n'}(q) \chi_{j \bar n}^{PF}(\bar q)
\chi^{U(1)}_{\bar n}(\bar q)\cr
=& \sum_j \sum_{ n +n'=0 mod 2 k_1 \atop n-n' =0 mod 2 k_2}
 \chi_{j n}^{PF}(q) \chi^{U(1)}_{n'}(q) \chi_{j }^{SU(2)}
 (\bar q).
 }}
We can check that it is modular invariant. Note that in the case
of $k_1=2$ it gives the $SO(3)$ model (depending on whether $k_2$
is even or odd we get the two $SO(3)$ modular invariants from
\part). If $k_1=1$ we recover the $SU(2)$ partition
function.

We can think of $(n-n')/2$ as the winding number $l_w$ which is
defined $\mod k_1$ through  $n-n' = 2 l_w k_2$, $l_w =0,1,..
k_1-1$. We could also define a ``momentum'' through $n+n' = 2l_p
k_1$  and $l_p$ is defined $\mod k_2$. We see that the model is
invariant under the exchange of $k_1$ and $k_2$ which amounts to
a reversal in the sign of $n'$ (or a T-duality in the circle
direction). It is interesting to note that, even though the
momentum and the winding are not conserved, the charge $n'= (l_p
k_1 - l_w k_2)$ is indeed conserved since it is the $U(1)_L$
charge which is left unbroken by the orbifold.

We see that \part\ is invariant under interchanging $k_1
\leftrightarrow k_2$, up to a T-duality that reverses the sign of
$n'$. Therefore, $ SU(2)_{k_1k_2}/\IZ_{k_1}=
SU(2)_{k_1k_2}/\IZ_{k_2}$.  In particular $SU(2)_k/\IZ_k =
SU(2)_k$.

Note that this theory, given by \part , is an asymmetric orbifold.
So D-branes in this theory are  examples of D-branes in asymmetric
orbifolds.

We note that all these theories including the $SU(2)_k$ theory
are rational with respect to the chiral algebra
$\CC=\CA^{PF_k}\times U(1)_k \subset SU(2)_k$.  The diagonal
modular invariant of this algebra is $\sum_{jnn'}
|\chi_{jn}^{PF}|^2|\psi_{n'}^{U(1)}|^2$.  Other partition
functions like the diagonal modular invariant of $SU(2)_k$ or
\part\ are different modular invariants of $\CC$. In particular,
the chiral algebra of $SU(2)_k$ can be thought of as the extended
chiral algebra of $\CC$ which can be denoted as $\CC/\IZ_k$.
Below, we will find this interpretation of the $SU(2)$ theory
convenient.  All the branes we will consider can be found using
the Cardy procedure \cardy\ and its generalization in
\refs{\fs,\bfs,\behrend} with respect to the algebra $\CC$.  In
doing that one must make sure that the expansion of the boundary
states in terms of Ishibashi states includes only closed string
states which are present in the theory.  Below, we do not follow
this algebraic route. Rather, we will  construct the branes using
a more geometric procedure.

\subsec{Branes in  $SU(2)_k$ and $SU(2)_k/\IZ_k $}

The standard Cardy theory applied to the $SU(2)$ level $k$ WZW
model produces a set of Cardy states
 \eqn\dzcardy{|A , \hat
j\rangle_C=\sum_{j}{S_{\hat j}^{~j}\over \sqrt{S_{0}^{~j}}}|A,j
 \rangle\rangle.}
These states have been  interpreted geometrically as states made
out of $2\hat j+1$ D0-branes. These grow into D2-branes which
wrap conjugacy classes in the group
\refs{\ars\bds\as-\fredenhagen}.

It is convenient to express the $SU(2)$ Ishibashi states as
\eqn\ishicos{|A,j\rangle\rangle =\sum_{n=1
}^{2k}{1+(-1)^{2j+n}\over 2} |A, j, n\rangle\rangle_{PF}
|A,n\rangle\rangle_{U(1)}}

When we quotient the theory by $\IZ_{k_1}$ we find that not all
$SU(2)$ Ishibashi states appear in the quotient theory. More to
the point, the Cardy states \dzcardy\ are not invariant under
$\IZ_{k_1}$. We can form $\IZ_{k_1}$ invariant states by adding
all $\IZ_{k_1}$ images. This automatically projects onto
Ishibashi states that are present in the quotient theory and thus
produces Cardy states that exist in the Lens space theory.

We now follow the strategy which proved useful above.  We consider
the case $k_1 = k$ and perform T-duality to the original
$SU(2)_k$ theory.  Then we reinterpret these branes as new
B-branes in $SU(2)_k$. We find
 \eqn\newcardys{\eqalign{|B,\hat j,\eta=\pm 1\rangle_C &=
 \sum_{j=0}^{k/2}{\sqrt k S_{\hat j}^{~j}\over \sqrt{S_{0j}}}
 \left[ {1+(-1)^{2j}\over 2}|Aj0\rangle\rangle_{PF}|B00
 \rangle\rangle_{U(1)} \right. +\cr
 &\qquad  \left. \eta (-1)^{2 \hat j} {1+(-1)^{2j+k}\over 2}|A j k
  \rangle\rangle_{PF}|B,k,-k\rangle\rangle_{U(1)} \right] \cr
 & =\sum_{j=0 \atop j \in \IZ }^{k/2}{\sqrt k S_{\hat j}^{~j}\over
 \sqrt{S_{0j}}} |Aj0\rangle\rangle_{PF} \left[|B00
 \rangle\rangle_{U(1)} + \eta |B,k,-k\rangle\rangle_{U(1)}
 \right]\cr
 &={1\over \sqrt k} |B\eta\rangle_C^{U(1)} \sum_{\hat n=0}^{2k-1}
 |A \hat j \hat n \rangle_C^{PF}.}}
Note the use of the two B-branes in $U(1)_k$ defined in
\uocardyb. The last line of \newcardys\ shows that $|B,\hat
j,\eta\rangle_C= |B,{k\over 2}- \hat j,\eta \rangle_C$, and
therefore the different branes are labeled by $\hat
j=0,\half,...,\half [{k\over 2}]$.

As we mentioned above, we can view the $SU(2)_k$ model as an
orbifold $SU(2)_k = ( U(1)_k \times PF_k )/\IZ_k$. The last line
of \newcardys\ becomes easy to understand, we just take products
of Cardy states and we take a superposition that is invariant
under $\IZ_k$.  Similarly, we can express \dzcardy\ as
 \eqn\acardydi{|A,\hat j\rangle_C={1\over \sqrt k} \sum_{\hat n}
 |A\hat j\hat n\rangle_C^{PF}|A\hat n\rangle_C^{U(1)}}
which is obvious from the orbifold interpretation but can also be
checked explicitly using the expressions for the various factors.
It is now straightforward to compute the annulus diagrams
 \eqn\annuls{\eqalign{
&{}_C\langle B,\hat j\eta|q_c^{L_0-{c\over 24}}|B,\hat
j'\eta'\rangle_C= \sum_{j} \sum_{n=0}^{2k -1}
\sum_{n'=0}^{2k-1} N_{\hat j\hat j'}^{j}\chi_{j,n}(q_o)
\psi_{n'}(q_o) {1+\eta\eta' (-1)^{n'}\over 2} \cr
 &{}_C\langle A\hat j|q_c^{L_0-{c\over
24}}|B\hat j'\rangle_C=\sum_{j,n}N_{\hat j\hat
j'}^{j}\chi_{j,n} (q_o){(q_o)^{1\over 48} \over
\prod_m(1-(q_o)^{m+\half} )}.}}

The geometrical interpretation of the B-type state is that it
represents a 3-dimensional object corresponding to a
``thickened'' or ``blown-up'' $D1$ string. This can be
substantiated by the computation of the shape of the brane at
large $k$ (see appendix D) and from the various properties listed
in the following comments.

\item{1.}
  From the mode expansion in \annuls\ we can read off the
spectrum of the open strings stretched between two B-branes or
between a B-brane and the familiar A-branes.
\item{2.} The normalizations in \newcardys\  can be deduced from the
T-duality argument  and is  such that the coefficients of the
various modes in \annuls\ are nonnegative integers.
\item{3.}
The label $\eta$ in the state can be interpreted as a Wilson
line around the B brane.  To see that, note that a Wilson line
introduces a phase $e^{i\pi a n/k}$ in front of a boundary state
$|B,n,-n\rangle\rangle_{U(1)}$. For the special values $a=1$ and
$n=k$ this corresponds to $\eta=-1$ in \newcardys. For generic
values of $a$ the boundary state breaks the extended $U(1)_k$
chiral algebra because a state with momentum $n$ gets a
phase different from a state with momentum $n+2k$. Other values of $a$ are
of course good Cardy states, and it is straightforward to describe
them.  We have to use Ishibashi states and characters of the
noncompact $U(1)$ generated by the current instead of those of
$U(1)_k$.  This comment follows from the behavior of the circle
and is identical to the phenomena we saw in section 3.1.
\item{4.} The various D-brane charges of $|B\hat j\pm\rangle$ can be
determined by the overlap with $|A\hat j=\half,\hat n=
0\rangle\rangle$.  Since this overlap vanishes, all these charges
also vanish.  Note that this is unlike the situation with the
A-branes $|A \hat j\rangle$, which have nonzero D0-brane charge.
We should comment, however, that the issue of the charge makes
sense only in the context of the superstring.
\item{5.} The mass of $|B\hat j\eta\rangle$ is determined by the
overlap with $|A, j=0, n=0\rangle\rangle$.  We easily find
\eqn\massdone{M(B\hat j)=\sqrt k M(A\hat  j) \sim k^{3\over 2 }
\sin{\pi (2\hat j+1)\over k+2}= \cases{ k^{3\over 2}\sin{2\pi
\hat j\over k} & $\hat j \approx {k \over 4}  \gg 1$ \cr
 k^{1\over 2}(2\hat j+1) & $k\gg \hat j \approx 1 $.}}
This fact has the following interpretation.  The radius of the
target space is $\sqrt k$. Therefore the new branes have the same
width as the corresponding A-branes but they wrap the target
space.  They have the form of a disk (as in section 4.2) times a
circle of the size of the group.
\item{6.} The factor ${(q_o)^{1\over 48} \over \prod_m(1-(q_o)^
{m+\half})}$ in the second inner product in \annuls\ can be
interpreted, as in flat space as due to a single boson with D
boundary conditions at the D0-brane and N boundary conditions at
the D1-brane.
\item{7.}  Notice that since the T-duality that maps $SU(2)_k/
\IZ_k$ to $SU(2)_k$ only changes the $U(1)$ part the states
continue to obey A-type boundary conditions for the parafermion
part but obey B-type for $U(1)$.
\item{8.} We could also have taken a B state for the parafermions
and an A state for the $U(1)$. This would have given states
that differ from \newcardys\ by a rotation in the group.
\item{9.}  The diagonal matrix elements ${}_C\langle B,\hat
j={k\over 4}, \eta | q_c^{L_0-{c\over 24}}|B,\hat j={k\over 4},
\eta \rangle_C$ \annuls\ show that the open string Hilbert spaces
on these D-branes include the identity operator twice. Therefore,
as in the parafermion theory, the B-branes with $\hat j={k\over
4}$ are reducible and can be written as sums of two branes.  We
expect, by analogy with the parafermion theory, that these
D-branes are stable. Also, we expect them to cover the whole
$S^3$ target manifold but to have singularities.  We did not
explore in detail the properties of these branes to verify these
expectations. In order to think about these special branes it is
convenient to think of $SU(2)_k = (U(1)_k \times PF_k )/\IZ_k $,
and construct Cardy states that are products of B branes in PF
and A branes in U(1). If we take the special branes \newirred ,
multiply by $U(1)$ A branes and take all the $\IZ_k$ images, we
get the irreducible branes at $\hat j =k/4$ that we mentioned
above.

In order to explore the shape of the B-branes we consider the
limit $k\gg \hat j \gg 1$ and examine the open strings which live
on them. From \annuls\ we see that the spectrum is
\eqn\deltajnnp{\Delta_{j n n'}\approx{j(j+1)\over k} - {(n)^2\over
4k}+ {(n')^2\over 4k}\qquad j\in \IZ; \quad n=-2j,-2j+2,...,2j;
 \quad n'\in 2\IZ.}
The quantum number $n'$ is easily interpreted as momentum along a
circle.  $j'$ and $n$ give us information about the transverse
shape of the brane, $n$ can be thought of as angular quantum
number around the D1. The simplest case is when $\hat j =0$. This
corresponds to a D1-brane along a maximum circle  of $S^3$. It is
unstable and the instability corresponds to the D1-brane
shrinking. The instability can be seen in \annuls\ and comes from
the term with $j =0=n'$, $n =\pm 2$. The conformal weight of this
state is $\Delta = 1 - {1 \over k}$, this agrees with the
semiclassical expectations for such a D1-brane. It also has the
expected angular momentum around the D1.  More generally, this is
exactly the spectrum discussed on the D2-brane in section 4.2
tensored with a spectrum of open strings on a circle.

We expect to find similar B branes in the $SL(2)$ WZW model. In
fact the D1 in that case would describe D-particles moving in
$AdS_3$. The A-branes in $SL(2)$ are those discussed in
\refs{\stanciu\fstanciu\bachas-\hirosi}. These B branes in $SL(2)$
would describe D0 branes at the tip of  the cigar. It would be
interesting to work this out explicitly. It is interesting to note
that the B branes of Liouville theory might be related to the
construction  in \zam  , which seems to be somehow localized in
the Liouville direction.

Some classical D-string surfaces that seem to look like our
B-branes in $SU(2)$ were discussed in the classical analysis of
\stanciu , based on the idea of flipping the sign in the $SU(2)$
currents $J^a = - \bar J^a$. This boundary condition is not
consistent quantum mechanically because it does not respect the
{\it bulk } $SU(2)$ commutation relations (OPEs) for the current
algebra (independently of whether they they do or do not preserve
the symmetry)\foot{ We thank V. Schomerus for a discussion on
this point.}. Our construction does not translate into a simple
boundary condition for the currents. Qualitatively one can think
about the boundary conditions in the following way. We can write
the $SU(2)$ currents as $J^+ = e^{+ i \phi} \psi^+ $ and $J^- =
e^{- i \phi} \psi^- $. If we  think of $\psi^\pm$ as the
elementary parafermion fields then we are imposing the boundary
conditions $ \partial \phi = -\bar \partial~ \bar \phi$ and
$\psi^\pm =  \bar \psi^\pm$. So we flip the sign of the $U(1)$
boson but we do not do anything to the parafermions. This
description is not precisely  correct since $\psi^\pm$ are not
holomorphic fields in the theory, the proper statement is that
the appropriate powers of $\psi^\pm$ that do form holomorphic
fields are reflected without any  change. This preserves the
$SU(2)$ current OPEs.

Classical analysis of D-branes in $SU(2)$ appeared also in
\klimcik. The precise relation of their results to ours is
not clear to us.

\subsec{The $U(1)$ symmetries of the B-branes in $SU(2)$}

The B-branes in $SU(2)$ can be thought of as fat D1-branes. It helps
to consider the case of $\hat j$ small to think about this.  In
the parametrization
 \eqn\coord{ z_1 = e^{ i  \phi} \sin \theta ~,~~~~ z_2
 = e^{ i \tilde  \phi} \cos \theta ~,~~~~ds^2 = d\theta^2 + \sin^2
 \theta d { \phi}^2 +\cos^2 \theta d {\tilde \phi}^2 }
the D1 brane is at $z_1 =0$.\foot{ This brane actually differs
from \newcardys\  by  a trivial  $\pi$ rotation.} Geometrically
it appears to preserve a $U(1)_L\times U(1)_R$ subgroup of the $
SU(2)_L\times SU(2)_R$ isometry of the $S^3$ target space. In the
CFT we only saw a single $U(1)$ (and a $\IZ_k$). How did this
happen?

The $U(1)_L \times U(1)_R$ symmetries are generated by
$J^3_{L,R}$. In the B  boundary state the combination generated by
$J_L + J_R$ is obviously conserved, while $J^3_L - J^3_R$ is
conserved modulo $k$. More geometrically, in the coordinate
system \coord\ the brane is concentrated at $ z_1=\theta = 0 $.
The operator $J_L + J_R$ generates a shift of $ \tilde \phi$ and
corresponds to translations along the brane. It leads to a
conserved charge.  The operator $J_L - J_R$ generates a shift of
$\phi$ and corresponds to rotations around the brane. It does not
lead to a conserved charge.  The violation of this charge can be
seen in the boundary state \newcardys.  The term
$|B,k,-k\rangle\rangle_{U(1)}$ in the boundary state shows that
the brane can absorb a closed string state with $|j,k,-k\rangle$.
Therefore, $J^3_L + J^3_R$ is conserved while $J^3_L-J^3_R$ is
violated by $k$ units.

To see this more explicitly we consider a point like closed
string at $\theta = \pi/2 $ moving along $ \phi$. Due to the giant
graviton effect \suss\ it can  become a string that winds along
the $\tilde \phi $ angle. Let us see whether this is possible.
The action for a string with $t = \tau$ and $  \tilde \phi =
\sigma$ is
 \eqn\action{S = - \int dt \left[\sqrt{k} \cos \theta
 \sqrt{ 1 - k \sin^2 \theta{ \dot { \phi}}^2 } - k \cos^2
 \theta \dot{  \phi}\right], }
where the linear term in $\dot { \phi}$ comes from the $B$
field. The conserved momentum conjugate to $\dot{ \phi}$ is
 \eqn\conjug{J = { \sqrt{k} \cos \theta k \sin^2 \theta \dot
 { \phi}  \over\sqrt{ 1 - k \sin^2 \theta { \dot {
 \phi}}^2 } } +  k \cos^2 \theta,}
and therefore the energy can be expressed in terms of $J$
\eqn\energy{
E = \sqrt{k} \sqrt{ {( 1-{J\over k})^2 \over \sin^2\theta}  + (2
 {J\over k} -1)}  }
It is easy to check that when $ J < k $ the minimum energy is
obtained for $\sin \theta = 1$ and is $E = J/\sqrt{k}$, while for
$J =k$ we obtain $E = \sqrt{k}$ independently of the value of
$\sin\theta$. So we can continuously change between a ``momentum''
mode at $\theta =\pi/2$, and a winding mode, at $\theta = 0$.
Indeed at $\theta = 0$ and $J =k$ the string coincides with the
brane and it can, therefore, be absorbed by it.

An alternative way to see the violation of the $U(1)$ symmetry is
through an instanton calculation, which is essentially the
Euclidean version of the previous calculation.  The relevant
instanton is a disk instanton, whose boundary is a D1-brane at
$z_1=0$ (in the coordinate system \coord).  We can parametrize
the worldsheet by $|z_2| \le 1$, and choose a worldsheet such that
the phase $ \phi$ of $z_1=e^{i \phi}\cos \theta $ is
constant.

In order to calculate the instanton action we need to find a
convenient form of the $B$ field.  The $H$ field in the
coordinate \coord\ is given by the volume form $H={k\over
2\pi}\sin (2\theta) d\tilde \phi d \phi d\theta$.  A possible
choice of the $B$ field is $B= {k\over 2\pi} \phi \sin (2\theta)
d\tilde \phi d\theta $. Therefore, the WZ term of the disk
instanton is
 \eqn\wzterm{i\int B=ik \phi.}
Clearly, the $U(1)$ symmetry of shifting $\phi$ by a
constant is broken to $\IZ_k$.

\subsec{The moduli space of the B-branes in $SU(2)$}

With this understanding of the symmetries we can determine the
moduli space of the B-branes. The group that acts on the branes
is the isometry $SU(2)\times SU(2)$. The geometric picture
suggests that a $U(1)_\phi\times U(1)_{\tilde \phi}$ subgroup of
it, which shifts $\phi$ and $\tilde \phi$ (in \coord) by
constants leaves the branes invariant.  Therefore we might be
tempted to guess that the moduli space of these branes is
$SU(2)\times SU(2) \over U(1)_\phi\times U(1)_{\tilde \phi}$.
This cannot be the right answer because, as we explained above
$U(1)_{ \phi}$ is broken to $\IZ_k$, and is not a symmetry
of the problem. Therefore, it cannot appear in the denominator of
the quotient.

The resolution of the problem is easy when we remember that the
B-branes also have another modulus.  Since these branes are fat
D1-branes, we can turn on Wilson lines along them.  These take
values in $U(1)$ (we have already encountered some discrete
values of these Wilson lines in the parameter $\eta$ in
\newcardys) -- we add  $\oint A= n\rho$ to the worldsheet action
where $n$ is the number of times the boundary of the worldsheet
winds around the D-brane.  Now \wzterm\ is modified to
\eqn\wzterm{i\int B+i\oint A =ik\tilde \phi + i\rho.}
The $U(1)_{\tilde \phi}$ symmetry is now restored, if at the same
time we also shift the value of the Wilson line $\rho$.  We
conclude that the moduli space of these D-branes is
\eqn\modulif{{SU(2)\times SU(2) \times U(1)_W\over U(1)_\phi
\times U(1)_{\tilde \phi}},}
where  $U(1)_W$ is the value of the Wilson line and $U(1)_{
\phi}$   not only shifts $ \phi$ (i.e.\ is in $SU(2)\times
SU(2)$) but also shifts $\rho$ (i.e.\ is in $U(1)_W$).

As a simple consistency check of the answer \modulif, we note
that the spectrum of open strings on a B-brane can be read off
from \annuls.  It includes five dimension one operators which are
moduli.  One of them is the descendant of the identity operator
-- the $U(1)$ current.  The other four are the parafermion
primaries $(j=0,n=\pm 2)$ with $U(1)$ charges $n'=\pm 2$ for the
compact boson.

\subsec{D-branes in more general Lens spaces}

In this subsection we explore some of the D-branes in the Lens
spaces $S^3/\IZ_{k_1}^L$.

The A-branes are easily found by starting with the A-branes in
the $SU(2)_k$ covering theory.  In terms of the Cardy states in
the covering theory we have to sum over images.  These images are
rotated in the group and are not simply of the form $|A,\hat
j\rangle_C$ we used above.  However, in terms of Ishibashi states
the $\IZ_{k_1}^L$ projection is easily implemented.  It states
that in $|A,j,n\rangle\rangle_{PF}|A,n,n\rangle\rangle_{U(1)}$ the
value of $n$ must satisfy $n=0\mod k_1$.   Therefore, we are led
to study the Cardy states
 \eqn\dbranes{ |A \hat j \rangle_C=
\sqrt{k_1} \sum_{j \atop n=0\mod k_1} {S_{\hat j  j}\over
\sqrt{S_{0j} }} {1+(-1)^{2j+n}\over 2}|A j, n \rangle\rangle_{PF}
|A,n,n \rangle\rangle_{U(1)}.
 }
More general A-branes can be found by rotating these branes in
the target space $S^3/\IZ_{k_1}^L$.  For simplicity we will not
consider them.

The open string spectrum between these branes is easily computed
\eqn\openspe{ {}_C\langle A \hat j |q^{L_0-{c\over 24} }| A \hat
j' \rangle_C = \sum_{j\atop n-n'=0\mod 2k_2} N_{\hat j \hat j'}^j
 \chi^{PF}_{jn} \chi^{U(1)}_{n'}. }
We see that the spectrum includes the usual strings on a single
brane, as well as the strings stretched between the different
images. From this formula we see that all values of $\hat j$,
with $0 \leq \hat j < {k\over 4}$ give distinct branes, since the
open string spectrum on them is different. Furthermore, the
identity operator appears only once in the open string spectrum
on each brane, so these are irreducible branes. These branes can,
however, (even in the superstring) decay to each other. For $k$
even the brane with $\hat j = {k\over 4}$ is special because the
identity operator can appear twice on it. This happens if and
only if $k_1$ is even, as in $SO(3)$ ($k_1 =2$) \fs . In this
case this brane is reducible and should be expressed as a sum of
two other branes. It would be nice to explore this case in detail.

The theory also has B-branes
  \eqn\dthree{ |B \hat j \hat n \rangle_C= \sqrt{k_2} \sum_{j\atop
  n=0 \mod k_2 } e^{ {i\pi \hat n n \over k}} {S_{\hat j  j }
  \over \sqrt{S_{0j} }}{1+(-1)^{2j+n}\over 2} |A, j, n
  \rangle\rangle_{PF} |B,n,-n\rangle\rangle_{U(1)}. }
These are the B branes of the WZW which are invariant under the
$\IZ_{k_1}$ orbifold group. The parameter $\hat n$ is interpreted
as a Wilson line which takes $2k_1$ possible values.
 More precisely, as in the rational torus, the
Wilson line can have continuous values but our use of the full
$U(1)_k$ chiral algebra led us to only $2k_1$ values.  As for the
B branes in $SU(2)$, also here there exists nontrivial mixing
between rotations around the branes and the value of the Wilson
line.

We can calculate the open string spectrum between these branes:
 \eqn\dthreeopen{\eqalign{
 &{}_C\langle B \hat j \hat n|q^{L_0-{c\over 24} }| B \hat j'
 \hat n' \rangle_C =\sum_{j \atop n' - n = (\hat n -\hat n') \mod
 2k_1 } N_{\hat j \hat j'}^j \chi^{PF}_{jn} \chi^{U(1)}_{n'}\cr
 &{}_C\langle A \hat j|q^{L_0-{c\over 24} }| B \hat j' \rangle_C
 =\sum_{j n} N_{\hat j \hat j'}^j \chi^{PF}_{ jn} \chi^{U(1)}_{ND}.}}
Again we see that $\hat n$ is a Wilson line since it changes the
momentum quantization condition.

Similarly to the above discussion, for $k_2$ even the B branes
with $ \hat j ={k\over 4}$ are reducible and should be expressed
as sums of two separate branes.

These B branes on lens spaces seem relevant for producing the
theories described in \arched , since they live on a space with
an $H$ field with one index along their worldvolume.

\bigskip
\centerline{\bf Acknowledgements}

This work was supported in part by DOE grants \#DE-FG02-90ER40542,
\#DE-FGO2-91ER40654, \#DE-FG02-96ER40949, NSF grants PHY-9513835,
PHY99-07949, the Sloan Foundation and the David and Lucile Packard
Foundation. GM would like to thank the ITP at Santa Barbara for
hospitality during the writing of this manuscript. We would also
like to thank I. Brunner, M. Douglas, S. Giddings. J. Harvey. J.
Polchinski, G. Segal, S.  Shenker, and E. Verlinde, for
discussions.


\appendix{A}{Characters and modular transformations}

Here we collect relevant character and modular transformation
formulae for easy reference.

\subsec{Level $k$ theta functions}

We define the $\theta$ functions as \eqn\deftheta{
\Theta_{m,k}(\tau, z) \equiv \sum_{\ell\in \IZ  } e^{2\pi i \tau
k(\ell + {m\over 2k })^2 } e^{2\pi i z k (\ell + {m\over 2k })} }
Under modular transformations they transform as
 \eqn\modtrans{
\Theta_{m,k}(\tau, z) = { 1 \over \sqrt{\tau/i} } q'^{ k z^2/4}
{1\over \sqrt{ 2 k}} \sum_{m'} e^{ - i \pi m m'/ k}
\Theta_{m',k}(-1/\tau ,- z/\tau ) }
where $q' \equiv e^{ - 2 \pi i/\tau } $

\subsec{$SU(2)$ characters}

The $SU(2)$ characters are given by
 \eqn\charactersutwo{\eqalign{
 &\chi^{su(2)}( \tau, z ) \equiv
  q^{-c/24} Tr_j[ q^{L_0} e^{ i 2 \pi z J_0^3} ] =\cr
 &=  q^{ -c\over 24} { \sum_m
q^{ {1 \over k+ 2} [ (j +{1\over 2} + m (k+2) )^2 -1/4 ] }\left(
e^{i 2 \pi z (  j+ {1\over 2} + m(k+2) ) } - e^{-i 2 \pi z (  j+
{1\over 2} + m(k+2) ) } \right) \over ( e^{ i \pi z} - e^{-i \pi
z } ) \prod ( 1- e^{ i 2 \pi z} q^n) ( 1- e^{- i 2 \pi z} q^n)
 ( 1-  q^n)  }}}

In the above form we can see that the denominators come from the
three currents and the numerator takes into account the presence
of null states.

This character can also be written as a ratio of theta functions
\eqn\sutwochar{ \chi_{j}(\tau,z) = { \Theta_{(2 j+1),k+2} -
\Theta_{-(2j+1),k+2} \over \Theta_{1,2} - \Theta_{-1,2} } ~.}
We will use the formula
 \eqn\modtrans{ \chi^{su(2)}_j(\tau, z) =e^{-i 2 \pi k z^2/(4\tau)
  } S_{j}^{~j'} \chi^{su(2)}_{j'}(-1/\tau, - z/\tau) }

\appendix{B} {Computation of overlaps between A and B branes in
the parafermion theory}

We begin the computation by computing overlaps of A-type and
B-type Ishibashi states. We act on \btypepf\ with
$q_c^{H^{SU(2)}}$ where $H = L_0 - c/24$ and take an inner
product with an A-type Ishibashi state. Using
\eqn\youoneover{
\langle \langle A, n,n\vert q_c^{H^{U(1)}}\vert B, r,-r\rangle
\rangle = \delta_{n,0}\delta_{r,0}\chi_{ND}(q_c)
}
where
\eqn\defchnd{ \chi_{ND}(q_c) =  {1\over q_c^{1/24} \prod(1+q_c^n)}
}
 we obtain
 \eqn\overlap{\langle \langle A j 0 |q_c^{L_0-{c\over 24}}|B j'0
 \rangle \rangle= \delta_{j,j'} \chi'_j }
Here $\chi'_j$ is defined in terms of  $SU(2)$ characters as
 \eqn\newchar{\chi'_j = { \chi_j^{su(2)}( \tau, z ={1\over 2})
 \over \chi_{ND} }
 }
and  we are thinking of the SU(2) Ishibashi state as a product of
a parafermion state and a $U(1)$ state. Here $\chi^{su(2)}( \tau,
z )$ is the $SU(2)$ character given in \charactersutwo. ( Notice
that $\chi_j(\tau, z ={1\over 2})$ vanishes for half integer $j$.)

We will need to use the modular transformation properties of these
characters. Let us start with the modular transformation property
of the single boson character with ND boundary conditions. It has
a simple form
 \eqn\modchar{\chi_{ND}(q_c) ={1 \over  q_c^{1/24}
 \prod_{n=1}^\infty (1+
 q_c^n) } = {\eta(\tau_c)\over \eta(2\tau_c)} =
\sqrt{2}{\eta(\tau_o)\over \eta(\half \tau_o)} = \sqrt{2}
 {q_o^{ 1 \over 48} \over  \prod_{n=1}^\infty (1-
 q_o^{n-{1\over 2}}) } }
here $q_c=e^{2\pi i\tau_c}$ is the closed string modular parameter
and $q_o=e^{-2\pi i/\tau_c}=e^{2\pi i\tau_o}$ the open string one.
The formula \modchar\ can be understood simply by thinking of
$2/\chi_{ND}^2(q_c)$ as the partition function of two fermions
with $+,-$ boundary conditions on a torus.

Now we  transform the $SU(2)$ characters using \modtrans. Setting
$z={1\over 2}$ we find, for the parafermion characters we are
interested in
 \eqn\modPFchar{ \chi'_{j}(q_c) = { 1 \over \sqrt{2}} S_{j,j'} \tilde
 \chi_{j'}(q_o) }
where $\tilde \chi_j$ is defined as
\eqn\newchar{ \tilde \chi_{j}(q_o)  =  q_o^{  k/16 }
 \chi_j^{SU(2)}(\tau', \tau'/2)q_o^{ {1\over 24} } q_o^{- 1/16}
 \prod_{n=1}^\infty (1-q_o^{n-{1\over 2}})
}

Notice that $\tilde \chi_{j} = \tilde \chi_{{k\over 2}-j}$. This
can be seen most easily from the the modular transformation
property and the fact that $\chi'_j$ vanishes for half integer
$j$.

Finally we can compute
 \eqn\overlaPFin{\eqalign{
 {}_C\langle  B \hat j' |q_c^{L_0-{c\over 24}}| A \hat j \hat m
 \rangle_C = & (2k)^{1\over 4} \sum_{j \in \IZ} { S_{\hat j'}^{~  j}
 {S^{PF}}_{\hat j \hat m}^{~~ j 0} \over \sqrt{ S_{0}^{~j}}
 \sqrt{ {S^{PF}}_{00}^{~~j0} }} \chi'(q_c) =  \sum_{j \in \IZ} {
 S_{\hat j', j} S_{\hat
 j j} S_{j,j''} \over S_{0j} } \tilde \chi_{j''}(q_c)\cr
 & = \sum_{j''}
 N_{\hat j', \hat j}^{j''} \tilde \chi_{j''}(q_o)\cr
 }}
where in the second line we used the relation between the
parafermion S-matrix and the $SU(2)$ S-matrix and the modular
transformation properties of $\chi'_j$. We see that indeed the
coefficients are all integers.

In order to compute the overlaps with the special B branes
with $\hat j = k/4$ (when $k$ is even) we need the overlap
\eqn\compute{ \langle \langle A {k\over 4}
{k\over 2}  |q_c^{L_0-{c\over 24}}| B {k\over 4}{k\over 2}
-{k\over 2} \rangle\rangle.}
This overlap can be computed in terms of the S-matrix
for the 1-point blocks on the torus.

Referring to the decomposition \btypepf\ we see that the
parafermionic state, which exists in the closed string spectrum,
multiplies a state in the $U(1)$ theory which has squareroot
branch cuts with representations with $r$ odd. We can cure this
problem by acting with the operator in the $SU(2)$ theory: 
\eqn\specop{
\Phi^{j=k/2}_{m=k/2} (z) = \Phi^{k/2}_{k}(z) \otimes e^{i {k\over
\sqrt{2k}}\phi}(z)
= 1 \otimes e^{i {k\over \sqrt{2k}}\phi}(z).
}
We then obtain
\eqn\onepoint{
\langle \langle j' \vert (1\otimes \tilde \Phi^{k/2}_{k/2}(\bar z))
q_c^{H^{SU(2)}}
(1\otimes e^{i \pi \tilde J^1_0}) \vert j \rangle \rangle
=\chi_{ND}\langle \langle A {k\over 4}
{k\over 2}  |q_c^{L_0-{c\over 24}}| B {k\over 4}{k\over 2}
-{k\over 2} \rangle\rangle
}
The left hand side is a conformal block for the one-point
function on the torus of the $j=k/2$ operator (``simple
current''). Analogs of the Verlinde formula for these blocks were
worked out in \refs{\mooreseiberg,\fss} and have been applied to
boundary conformal field theory in \refs{\fs,\behrend}. In this
way one can complete the computation of the overlap, but we have
not carried out the details.

Finally, let us consider the B-A type overlaps in the
superparafermionic theory. From the coset decomposition we obtain
the equality of Ishibashi states
\eqn\superishis{
\vert j \rangle\rangle \otimes \vert s \rangle\rangle
= \sum_{r=0}^{2k+3} \vert j,r,s\rangle\rangle \otimes
\vert r\rangle\rangle
}
The B-type states are now defined by
\eqn\superishisb{
(1\otimes e^{i \pi \tilde J^1_0})\vert j \rangle\rangle
\otimes \vert B s,-s \rangle\rangle
= \sum_{r=0}^{2k+3} \vert B,j,r,s;-r-s\rangle\rangle \otimes
\vert B, r,-r\rangle\rangle
}
from which we obtain
\eqn\overlapspb{
\langle \langle A j,n,s |q_c^{L_0-{c\over 24}}|B j',n',s'
 \rangle \rangle=\delta_{n,0}\delta_{n',0}\delta_{s,0}
 \delta_{s',0} \delta_{j,j'}
\chi^{SU(2)}_j(\tau_c,z=1/2)
 }
for $j\not= k/4$, with more elaborate formulae in this latter
case. In \overlapspb\ the delta functions should be understood up
to the spectral flow identification. The twisted-scalar function
$\chi_{ND}$ has cancelled from the LHS and RHS. {}From this
expression one easily checks the integrality of the coefficients
in the open string channel for A-B overlaps.

\appendix{C}{Explicit low $k$ examples of branes in bosonic
parafermions}

The purpose of this appendix is to demonstrate our general
formalism in special solvable examples at low values of $k$.

\subsec{$k=2$}

The simplest parafermion theory is \eqn\simplektwo{{SU(2)_2\over
U(1)_2},} which is the same as the Ising model.  Its fields are
easily identified as \eqn\isingfields{\matrix{ (j,m)& \Delta &
\cr (0,0) & 0 & {\bf 1} \cr (1,0) & {1\over 2} & \psi \cr
({1\over 2}, 1) & {1\over 16} & \sigma }}

The three Cardy states $|A, 0,0\rangle_C$, $|A, 1,0\rangle_C$ and
$|A, {1\over 2}, 1\rangle_C$ have masses proportional to $1$, $1$
and $\sqrt 2$ respectively. At small level one cannot necessarily
trust the geometrical picture. Nevertheless, it works remarkably
well. Geometrically, we can interpret the  first two states as
D0-branes at the two special points on the disk  and the third is
a D1-brane connecting them.  The spectrum of open string on these
D-branes is $\chi_{0,0}$, $\chi_{0,0}$ and $\chi_{0,0}+\chi_{1,0}$
respectively.

Our general discussion predicts three B-branes at the center of
the disk $|B,\hat j=0\rangle_C$, $|B,\hat j=\half,\eta=\pm 1
\rangle_C$, which are made out of a single or two D0-branes. Their
masses should be $\sqrt 2$, $1$ and $1$ respectively and with
spectrum of open strings on them $\chi_{0,0}+\chi_{1,0}$,
$\chi_{0,0}$ and $\chi_{0,0}$.  Since this is a discrete series
theory, it should not have any more Cardy states in addition to
the three A-branes mentioned above.  Indeed, it is easy to
identify  $|B,\hat j=0\rangle_C=|A,{1\over 2}, 1\rangle_C$ and
$|B,\hat j=\half ,\eta=\pm 1\rangle_C=|A ,\hat j=0, \hat m=
0,1\rangle_C$.

The fact that in this model the A-branes are the same as the
B-branes is a consequence of the self-duality of the Ising model.

\subsec{$k=3$}

The second simplest parafermion theory is
\eqn\simplep{{SU(2)_3\over U(1)_3}.} It has $c={4\over 5}$ and
representations: \eqn\spararep{\matrix{(j,n)& \Delta \cr
(0,0)\sim({3\over 2},3)\sim ({3\over 2},-3)  & 0 \cr (1,0)
\sim({1\over 2},3)\sim ({1\over 2},-3)& {2\over 5}\cr ({1\over
2},\pm1) \sim (1,\pm2) & {1\over 15}\cr ({3\over 2},\pm1) \sim
(0,\pm2) & {2\over 3}\cr}}

Since $c$ is less than one, this theory is also a member of the
Virasoro discrete series ($p=4$ when $c=1-{6\over p(p+1)}$), and
hence can be represented as the coset $SU(2)_3\times SU(2)_1
\over SU(2)_4$. The parafermion theory is not the standard modular
invariant of this theory but the one with an extended chiral
algebra. It can be denoted as $SU(2)_3\times SU(2)_1 \over
SO(3)_4$.  The representations of the discrete series are labeled
by two integers: $r=2j_1+1=1,...,4$ and $s=2j_2+1=1,...,5$
corresponding to the spins of two of the $SU(2)$ factors in the
coset (the spin of $SU(2)_1$ is determined by a selection rule)
and they are subject to the identification $(r,s) \sim
(5-r,6-s)$.  The dimensions of the representations are
\eqn\disdim{\Delta_{rs}={(6r-5s)^2-1\over 120}}
 leading to the spectrum of dimensions
\eqn\dimtab{\matrix{ s r &1&2&3&4\cr 1&0&{2\over 5}&{7\over 5}&
3\cr 2&{1\over8}&{1\over 40}&{21\over 40}&{13\over 8}\cr 3&{2\over
3}&{1\over 15}&{1\over 15}&{2\over 3}\cr 4&{13\over 8}&{21\over
40}&{1\over 40}&{1\over 8}\cr 5&3&{7\over 5}&{2\over 5}&0\cr}}

In terms of these representations the partition function of the
parafermion theory is
\eqn\paraparti{Z_{PF}=|\chi_{11}+\chi_{15}|^2+|\chi_{21}
+\chi_{25}|^2 +2|\chi_{13}|^2 +2|\chi_{23}|^2.} Note that as
expected from the coset with $SO(3)_4$ as opposed to $SU(2)_4$,
the chiral algebra was extended (by $\chi_{15}$) and
correspondingly, all representations with $s$ even (half integer
spin) were projected out.  Also, the two representations with
characters $\chi_{13}$ which are isomorphic as Virasoro
representations are different representations of the parafermion
algebra (similarly for $\chi_{23}$).  We will distinguish these
representations by denoting them by $13\pm$ and $23\pm$. These
representations can be identified with the parafermion
representations as \eqn\paramini{\eqalign{
&\chi_{11}+\chi_{15}=\chi_{j=0,n=0}\cr
&\chi_{13\pm}=\chi_{j=0,n=\pm 2}\cr
&\chi_{21}+\chi_{25}=\chi_{j=1,n=0}\cr
&\chi_{23\pm}=\chi_{j={1\over2},n=\pm 1}\cr}}

The Ishibashi states in the discrete series theory with the
standard modular invariant $\sum_{rs}|\chi_{rs}|^2$ are
determined by invariance under the Virasoro algebra.  They are
$|rs\rangle\rangle$ for all $r,s$ subject to the identification.

Let us examine the Ishibashi states in the parafermion theory.
Imposing invariance under the whole parafermion chiral algebra we
find
 \eqn\paraishi{\eqalign{ &|11\rangle\rangle+
|15\rangle\rangle=|A,\hat j=0,\hat n=0\rangle\rangle, \quad
|13\pm\rangle\rangle=|A,\hat j=0,\hat n=\pm2\rangle\rangle  \cr
&|21\rangle\rangle+|25 \rangle\rangle=|A,\hat j=1,\hat
n=0\rangle\rangle, \quad |23\pm\rangle\rangle=|A,\hat j={1\over
2},\hat n=\pm1\rangle\rangle.}} Out of these states we can form
Cardy states \eqn\paracardy{|A,\hat j=0,\hat n=0\rangle _C, \quad
|A,\hat j=0, \hat n=\pm2\rangle _C , \quad |A,\hat j=1,\hat
n=0\rangle _C, \quad |A,\hat j={1\over 2},\hat n=\pm1\rangle _C}
by using
 \eqn\ishiscarpar{|A, \hat j,\hat n\rangle
_C=\sum_{j'n'}{S^{PF}_{\hat j\hat n j'n'}\over
\sqrt{S^{PF}_{00j'n'}}}|A,j'n'\rangle\rangle, \qquad \qquad
S^{PF}_{\hat j\hat nj'n'}= {2\over \sqrt{15}}e^{i\pi \hat nn'\over
3}\sin{\pi(2\hat j+1)(2j'+1)\over 5}.}

The two Ishibashi states
 \eqn\twomoreis{|11\rangle\rangle-
|15\rangle\rangle, \quad |21\rangle\rangle - |25 \rangle\rangle}
satisfy all the consistency conditions including those arising
from the Virasoro algebra.  They are not invariant under the
parafermion chiral algebra but this is not a fundamental
consistency condition. It is clear that the eight dimensional
space spanned by \paraishi\ and \twomoreis\ is the entire set of
Ishibashi states in the parafermion theory.  With these Ishibashi
states we identify two new Cardy states
\eqn\newcardy{\eqalign{ &|B,\hat j=0\rangle _C={\sqrt 2 S_{1211}
\over \sqrt{S_{1111}}}(|11\rangle\rangle- |15\rangle\rangle)
+{\sqrt 2 S_{1221} \over \sqrt{S_{1121}}} (|21\rangle\rangle - |25
\rangle\rangle )\cr &|B,\hat j=1\rangle _C={\sqrt 2 S_{3211} \over
\sqrt{S_{1111}}}(|11\rangle\rangle- |15\rangle\rangle) +{\sqrt 2
S_{3221} \over \sqrt{S_{1121}}} (|21\rangle\rangle - |25
 \rangle\rangle ),\cr}}
where we have used the $S$ matrix of the minimal model
\eqn\sminimal{S_{rsr's'}={2 \over \sqrt{15}}
(-1)^{(r+s)(r'+s')}\sin({\pi rr'\over 5}) \sin ({\pi ss'\over
6}).}

Using this $S$ matrix we can find the open string spectrum between
the various D-branes by calculating the inner products
\eqn\innerpr{\eqalign{ &{}_C\langle B,\hat j=0|q_c^{L_0-{c\over
24}}|B,\hat j=0\rangle _C= \chi_{11} +\chi_{15}+2\chi_{13}=\cr
&\qquad \qquad \qquad \chi_{j=0,n=0}+\chi_{j=0,n=2}
+\chi_{j=0,n=-2}\cr
 &{}_C\langle B,\hat j=1|q_c^{L_0-{c\over
24}}|B,\hat j=1\rangle _C=\chi_{11}+\chi_{15}+2\chi_{13}
+\chi_{21}+\chi_{25}+2\chi_{23}=\cr
 &\qquad \qquad \qquad
\chi_{j=0,n=0}+\chi_{j=0,n=2}+\chi_{j=0,n=-2} +
\chi_{j=1,n=0}+\chi_{j={1\over 2},n=1}+\chi_{j={1\over 2},n=-1}\cr
& {}_C\langle B,\hat j=0|q_c^{L_0-{c\over 24}}|B,\hat j=1\rangle
_C= \chi_{21}+\chi_{25}+2\chi_{23}=\cr
 &\qquad \qquad \qquad
\chi_{j=1,n=0}+\chi_{j={1\over 2},n=1}+\chi_{j={1\over 2},n=-1}\cr
 &{}_C\langle B,\hat j=0|q_c^{L_0-{c\over 24}}|A, \hat j=0,\hat n
 \rangle{}_C= \chi_{12} +\chi_{14}\cr
  &{}_C\langle B,\hat j=0|q_c^{L_0-{c\over
24}}|A,\hat j={1\over 2},\hat n\rangle _C=\chi_{22}+\chi_{24}\cr
 &{}_C\langle B,\hat j=1|q_c^{L_0-{c\over 24}}|A,\hat j=0,\hat n\rangle
 =\chi_{22}+\chi_{24}\cr
 &{}_C\langle B,\hat j=1|q_c^{L_0-{c\over
24}}|A,\hat j={1\over 2},\hat n\rangle_C=\chi_{12}+\chi_{14}+
 \chi_{22}+\chi_{24}.\cr}}
Here we interpreted the factors of two in the fusion rules to
lead to an answer which is compatible with the global symmetries
of the parafermion theory.

Note that the characters described in appendix B, which appear
when we compute the overlaps between A and B states, are, in this
case, \eqn\ekthree{ \eqalign{ \chi'_0 = &\chi_0 - \chi_3 \cr
\chi'_1 =& - \chi_{2/5} + \chi_{7/5} \cr \tilde \chi_{0} =
&\tilde \chi_{3/2} = \chi_{1/8} + \chi_{ 13/8} \cr \tilde
\chi_{{1\over 2}} = & \tilde \chi_{1} = \chi_{1/40} +
\chi_{21/40}. }}

It is also worth remarking that the above boundary states are the
same as those found by \behrend\ by associating nontrivial
representations of minimal model fusion rule algebras to the
nondiagonal modular invariants of the $c<1$ series.  They are
also mentioned in \fs. The significance of this example is that
since this model is a discrete series model, here we know that
there are no other D-branes in the system. This will not be the
case in the next example.

\subsec{$k=4$}

The next parafermion theory is based on the coset
\eqn\secsimp{SU(2)_4\over U(1)_4} and has $c=1$.  This theory is
isomorphic to a boson $\phi$ on $S^1/\IZ_2$ with $R^2=6$.  The
spectrum of primary fields is \eqn\primakfou{\matrix{ (j,m) &
\CO& \Delta \cr (0,0) & {\bf 1}& 0 \cr ({1 \over 2}, \pm 1) &
\sigma_{1,2}& {1 \over 16} \cr (1,0) & e^{i2\phi/\sqrt 6} + e^{-
i2\phi/\sqrt 6}& {1 \over 3} \cr (1, 2) & e^{i\phi/\sqrt 6} +
e^{- i\phi/\sqrt 6}& {1 \over 12} \cr ({1\over 2}, \pm 3) &
\tau_{1,2} & {9 \over 16} \cr (0,4) & \partial \phi & 1 \cr
(0,\pm 2) & e^{i3\phi/\sqrt 6} \pm e^{- i3\phi/\sqrt 6} & {3\over
4} \cr}} Where $\CO$ is the operator in terms of the orbifold
theory. Here the chiral algebra $\CA$ includes the even (under
$\phi\to -\phi$) operators in the chiral algebra of the rational
torus; e.g.\ $e^{i\sqrt 6\phi} + e^{- i\sqrt 6\phi}$.  The fields
$\sigma_{1,2}$ and $\tau_{1,2}$ are twist fields located at the
singularities of the orbifold.

We easily construct 10 Ishibashi $|A,jm\rangle\rangle$ states
which are invariant under $\CA $ and combine them into 10 Cardy
states $|A,\hat j\hat m\rangle_C$.  These 10 Cardy states can be
viewed either as states in the orbifold $S^1/\IZ_2$ theory or in
the disk theory. The four states $|A,\hat j=0,\hat m=0,\pm
2,4\rangle_C$ correspond to D0-branes at the four special points
on the boundary of the disk. The four states $|A,\hat j={1\over
2},\hat m=\pm 1,\pm 3 \rangle_C$ correspond to D1-branes stretched
between adjacent points on the boundary.  The two states $|A,\hat
j=1 ,\hat m=0,2 \rangle_C$ correspond to D1-branes stretched
between points on the boundary which are not adjacent.

In terms of the orbifold theory these 10 states can be
interpreted as follows.  The four states $|A,\hat j=0,\hat m=0,\pm
2,4\rangle_C$ are D0-branes stuck at the orbifold singularities.
The four states $|A,\hat j={1\over 2},\hat m=\pm 1,\pm 3
\rangle_C$ are D1-branes wrapping the orbifold once with the four
allowed values of the Wilson lines.  The two other states
$|A,\hat j=1 ,\hat m=0,2 \rangle_C$ are two D0-brane states at
points along the interval $S^1/\IZ_2$ corresponding to angles $\pm
{\pi\over 3}$ and $\pm {2\pi \over 3}$ in the $S^1$ (the $\IZ_2$
identification is $\alpha \sim -\alpha$).  This interpretation is
consistent with the fact that the masses of these D-branes are
proportional to $1\over 2$, $\sqrt 3\over 2$ and $1$
respectively, and with the spectrum of open strings stretched
between them.

It is clear from the orbifold point of view that the two
D0-branes on the interval can move and are part of a moduli space
of D0-branes.  Indeed, the open strings living on these D-branes
include massless moduli. It is not clear what this moduli
corresponds to in the disk picture.

{}From the general parafermion analysis we expect also four more
D-brane states at the center of the disk $|B, \hat j=0,{1\over
2}\rangle_C$, $|B, \hat j=1, \eta=\pm 1\rangle_C$.  In the units
we used above their masses are $1$, $\sqrt 3$, $1$ and $1$.  We
identify them in the $S^1/\IZ_2$ picture as follows.  $|B, \hat
j=0\rangle_C$ is a D0-brane at the center of the orbifold (angle
$\pm {\pi \over 2}$ on the covering $S^1$), $|B, \hat j={1\over
2}\rangle_C$ is two D1-branes wrapping the orbifold with opposite
values of the Wilson line (recall that in this configuration the
Wilson line is a modulus) at the symmetric value of the Wilson
line, and $|B,\hat  j=1,\eta=1\rangle_C$ ($|B,\hat
j=1,\eta=-1\rangle_C$) is a D0-brane at $\pm {\pi \over 6}$ ($\pm
{5 \pi \over 6}$). This interpretation is supported by their
masses, the spectrum of open strings on them, and the spectrum of
open strings stretched between them. The massless open string on
$|B, j={1\over 2}\rangle_C$ is interpreted in the orbifold
picture as the arbitrary value of the Wilson line, and the
massless open string on $|B, j=1, \eta \rangle_C$ is interpreted
as the position of the D0-brane.

This example is special because some of the branes we found are
part of a continuum of D-branes.  The generic D-brane on the
moduli space is not invariant under such a large chiral algebra
and hence it is neither an A-brane nor a B-brane.  The first
signal of such a moduli space is the existence of massless open
strings on the branes.  Such massless states ($\Delta=1$) occur
also for larger values of $k$, but it is easy to see that the
three point function of such open string vertex operators are
nonzero.  Therefore, they are not moduli and the D-branes are
isolated.

\appendix{D} { The shape of the branes }

\subsec{Preliminaries}

In this appendix  we will determine the shape of the branes. We
will always consider large $k$, since only in that case we can
talk about the shape of the brane in a classical geometrical way.
We first do the calculation for $SU(2)$ as in \fffs\ and then we
proceed with the other cases. What we will do is to scatter a
massless closed string from the brane. \foot{The computation
which follows has some similarities to the computations of \fffs,
but in fact is different. Reference \fffs\  computes the change
in the metric and B-field induced by the presence of a brane in
the WZW model. The computation is analogous to measuring the
transverse metric of a flat brane in flat space $  h(r) \sim
const./r^{d_T-2} $ where $d_T$ is the number of  transverse
dimensions. }

This scattering amplitude is determined in terms of the overlap
 \eqn\defcalsh{\langle {\rm Boundary ~state |  closed ~ string}
 \rangle.}
We can think of the closed string state as a graviton with
polarizations in the other directions (directions not involving
the WZW model). In that case the wavefunction of the closed
string in the WZW directions involves only the primary states and
is characterized by $|{\rm closed ~ string}\rangle = | j, m, m';
0\rangle $, where the zero means that no descendant appears.
These states can be thought of as functions over the manifold for
$j \ll k$.

In this appendix we take $m=0,1/2,..$.  It is related to $n$ used
in the main text through $n =2m$.

When we scatter a closed string state we can pick any
wavefunction for it. A particularly useful function is a delta
function on the manifold. For the $SU(2)$ manifold we choose
$\delta (\vec \theta - \vec \theta' )$, where $\vec \theta$
denotes the three angles in some coordinate system.

Let us define some useful functions. The eigenfunctions of the
Laplacian on the group manifold can be written as
 \eqn\defofd{ {\cal
D}^j_{mm'} = \langle j m | g(\vec \theta ) | j m' \rangle
 ~,~~~~~\langle j m | j m'\rangle = \delta_{m,m'} }
where $|j,m\rangle$ are a basis for the spin $j$ representation of
$SU(2)$. We can determine the relation between the normalization
of $ {\cal D}^j_{mm'} $ and $| j,m,m'\rangle $ (which is
normalized to one) in the following way. First notice that
$|j,m\rangle $ in
\defofd\ is normalized to one. Then if we act with raising or
lowering operators of $SU(2)_L$ or $SU(2)_R$ then we would
produce the same factors on ${\cal D}$ as they produce on $|
j,m,m'\rangle $. Therefore, the relation between the two
normalizations is just a constant independent of $m,m'$. This
constant can be determined by taking the square of \defofd\ and
summing over $m'$ , so that we get $\sum_{m'} |\langle j m |
g(\vec \theta ) | j m' \rangle|^2 = 1$.  We find that
 \eqn\normal{
{\cal D}^j_{mm'} \sim { 1 \over \sqrt{ 2 j + 1 }} \langle \vec
 \theta | j,m,m'\rangle }
So that we can write a $\delta$ function as
 \eqn\deltafun{ \delta
(\vec \theta - \vec \theta') \sim \sum_{j,m,m'} (2 j + 1)
 {\cal D}^j_{mm'}(\vec \theta)^* {\cal D}^j_{mm'}( \vec \theta') \sim
\sum_{j,m,m'} { \sqrt{2 j + 1 }} {\cal D}^j_{mm'}(\vec
 \theta)^*\langle \vec \theta' | j,m,m'\rangle. }
Here the ranges of indices are  $|m|, |m'| \leq j$ and $j
=0,{1\over 2}, ...$.
We only want to consider closed string states which are well
localized. Therefore, we will really consider a smeared delta
function where states with $j\sim k$ are suppressed. We could do
this with a cutoff like $\exp[-j^2/\epsilon^2]$. We will let
$k\to \infty$ and then take $\epsilon\to 0$.

\subsec{Shape of A-branes in $SU(2)$ }

Now let us consider the boundary states in the $SU(2)$ WZW theory
which correspond to the usual conjugacy classes which we denote by
$| A,\hat j\rangle $.  Defining $\hat \psi \equiv \pi (2 \hat j +
1)/(k+2) $ we find
 \eqn\overlap{\eqalign{
{}_C\langle A,\hat j | \vec \theta\rangle \sim & \sum_{j m} { S_{\hat
j , j} \over \sqrt{ S_{0,j} } } \sqrt{ 2 j+1 } {\cal
D}^j_{mm}(\vec \theta)^* \cr & \sim {2^{1/4}\over \sqrt{\pi}}
(k+2)^{1/4}  \sum_j  \sin[( 2 j+
1 ) \hat \psi ] { \sin (2 j+1)\psi \over \sin \psi }\cr
& \sim {\sqrt{\pi}\over 2^{7/4}}{(k+2)^{1/4} \over \sin \hat \psi }\bigl(
 \delta ( \hat \psi - \psi) - \delta(\psi - \hat \psi + \pi )
 +\cdots\bigr) \cr
 }}
where we have used approximations valid for
$j \ll k$, e.g.  so that
$S_{0j} \sim (2 j+1)/(k+2)^{3\over 2} \sim (2 j+1)/k^{3\over 2}$.
This is justified by the smearing of the
delta function  as discussed above. In the periodic delta
function only the first term contributes because of the
restriction on the ranges of $\psi$.
 We also
used that the boundary state has the same $m$ for the left and
right movers and the fact that $ \sum_m {\cal D}^j_{mm} = {\sin
(2 j+1)\psi \over \sin \psi } $. Note also that $\psi$ is the
angle in the coordinates where the metric of the $S^3$  is
 \eqn\metric{
ds^2 = d\psi^2 + \sin^2 \psi d^2s_{S^2} } so we can think of a
point parametrized by $\psi$ as a rotation in $SU(2)$ by angle $2
\psi$.

We can see we get the expected result, saying that branes are
localized at the conjugacy class given by $\hat \psi $. The
factor of ${ 1 \over \sin \hat \psi}$ is also reasonable,
reflecting the fact that the tension of the D2-brane is getting
stronger of small $\hat \psi$ since $F$ is getting stronger. This
factor in the effective tension gives also the right mass after
integration over the $S^2$, $M \sim k  \sin \hat \psi $.

Note that in the above calculation  there is no restriction on the
size of $\hat \psi$, it could be very big and it could be a brane
that wraps the equator, or even branes that are close to $\hat
\psi \sim \pi$.  (Of course if the size of $\hat \psi $ is too
small then the approximate $\delta$ function we were considering
above would not resolve the brane appropriately).

\subsec{Shape of B-branes in $SU(2)$ }

Now let us consider the B branes defined in \newcardys. We will
disregard the second term in the large $k$ limit. When we
multiply by the delta  function state we find
 \eqn\deltadone{
\langle B, \hat j,\eta |\vec \theta \rangle \sim k \sum_j {\cal
D}^j_{00}  S_{\hat j j} }
 It is convenient to think about ${\cal D}^j_{00}$ in Euler angles
$\chi , \tilde \theta, \varphi$, see \metricsth .
Then we need to compute $ \langle 0 |
e^{ i\tilde  \theta J^1 } | 0\rangle $. These are
the Legendre polynomials
$P_j(\cos\tilde\theta) $ (note that only integer values of $j$ appear),
including the normalizations since $P_n(1) = 1$.

In order to compute the sum in \deltadone\ we should remember the
generating function formula for Legendre polynomials
 \eqn\gen{
 \sum_{n=0} t^n P_n(x) = { 1 \over \sqrt{ 1 - 2 t x + t^2 }}
} So we see that, with $x = \cos\tilde\theta$,
 \eqn\overapp{\eqalign{
\langle B \hat j |\tilde \theta \rangle \sim &
 - i e^{ i \hat \psi} \sum_{n=0}^{\infty} P_n(x) e^{ i n 2 \hat \psi }
 + c.c.  \sim
{ \Theta( \cos \tilde \theta -  \cos 2 \hat \psi)    \over \sqrt{ \cos
\tilde \theta - \cos 2 \hat \psi } } }} where $c.c. $
means complex
conjugate and $\Theta(z)$ is the usual step function which
vanishes when $z < 0$.

This same calculation can be applied to the parafermion case. One
finds a disk of radius $\rho = \sin (\pi (2\hj +1)/(k+2))$. It
can be checked that the $\tilde\theta$ dependent factor in
\overapp\ is the expected factor resulting from the open string
metric and dilaton in \openstr .

\subsec{ The shape of branes in the  parafermion theory}

Let us now carry out an analogous computation in the parafermion
theory. Referring to the geometry discussed in section 2.2 we see
that  wavefunctions on the disk are $SU(2)$  wavefunctions that
are invariant under translations of $\tilde \phi$. So the
wavefunctions of the parafermion theory are 
 \eqn\wavefunctions{ \Psi_{j,m}  = e^{ i \alpha_{j,m}}
\langle j, m | g | j, -m \rangle = e^{ i 2 m  \phi} \langle
j, m | e^{ i ( 2 \theta -  \pi ) J^1 } | j, m \rangle = { 1 \over
\sqrt{ 2j +1 }}\langle \theta,  \phi | j , m, m \rangle_{PF} }
where the last equality is just a statement of the normalization
of the wavefunctions and $\alpha_{jm}$ is a phase that is irrelevant
for our purposes.   Let us just remark that these functions
are not the usual spherical harmonics
 \eqn\spherical{
 Y_{l,m}(\tilde \theta, \varphi) = \langle l , 0 | g | l , m
 \rangle,
 }
but they are related to them. Of course for $m=0$ they are the
same as the Legendre polynomials we used earlier.

We can now build the delta function closed string state roughly as
before
 \eqn\deltaPF{ \delta( \vec \theta - \vec \theta' ) \sim
\sum_{jm} ( 2 j +1)  \Psi(\vec \theta)^*_{jm} \Psi(\vec
\theta')_{jm} =
 \sum_{jm} \sqrt{2 j+1} \Psi(\vec \theta)^*_{j,m} \langle \theta'
 | j, m ,-m \rangle }
where $|j, m ,-m\rangle$ is a closed string state normalized to
one.

The Cardy state is given in \cardyPF. Now in order to see what
state this is we should take the overlap with the delta function
state. We get
 \eqn\overlapp{\eqalign{
{}_C\langle A, \hat j, \hat m | \theta,  \phi  \rangle \sim
& \sum_{j} \sum_{m=-j}^j  S_{\hat j, j} e^{ i 4 \pi m \hat m \over
k } \Psi_{j,m}(\theta, \phi)^* \cr &=\sum_{j}
\sum_{m=-j}^j  \sin ( \hat \psi (2 j+1) )\Psi_{j,m}(\theta,
 \phi - 2 \pi \hat m/k)^* }}
 So we see that $\hat m$ is an
 angular position. Notice that we restricted the sum
over $m$ only to those states which can be thought of as
wavefunctions on the space. So we can set $\hat m$ =0 and get the
other states by acting with $\IZ_k$  rotations in $\phi $.
So we need to calculate
 \eqn\calcula{ \sum_{ m
=-j}^j  \langle j ,m | e^{ i 2  \theta J^1} | j, - m\rangle e^{ i
2m  \phi} =  \Tr[ e^{i 2  \phi J^3} e^{ i (2 \theta -
\pi) J^1 }  ]
 = { \sin ( 2 j + 1)\psi \over \sin \psi }
 }
where
 \eqn\valuepsi{ \cos \psi = \cos  \phi \sin\theta }
When we insert this back into \overlapp\ we find
 \eqn\finalshape{ {}_C\langle A,
\hat j, \hat m =0 | \theta,  \phi \rangle \sim { \delta(
\hat \psi - \psi(\theta, \phi) ) \over \sin \psi } }

Notice that this delta function defines a line in the two
dimensional manifold where the parafermion theory lives.  A
straight vertical line in polar coordinates obeys $ \rho \cos
\phi = \sin\theta \cos\phi = {\rm constant}$, which is indeed
what we find from \calcula \finalshape . Straight lines at other
angles are obtained by performing rotations, which amounts to
taking nonzero $\hat m$. In the super symmetric case we rotate
by $2 \pi \hat m/(k+2)$, and we must include the sector
from the $U(1)_2$ theory.

\appendix{E}{Some formulae from $\NN=2$ Representation Theory,
Minimal models, and all that}

The $U(1)$ generator of the $\NN=2$ algebra is normalized
so that
\eqn\ntwoalg{ \eqalign{ [J_n, J_m] & = {c\over 3} m
\delta_{n+m,0} \cr
J(z) J(w) & \sim {c/3 \over (z-w)^2} + \cdots  \cr} }
and in particular $[J_n, G_m^\pm] = \pm G_{n+m}^\pm$.

The spectral flow operation by $\eta$ is \schwimmerseiberg
\eqn\specflow{ \eqalign{ L_n & \to L_n + \eta J_n + {c\over 6}
\eta^2 \delta_{n,0} \cr J_n & \to J_n + \eta {c\over 3}
\delta_{n,0} \cr G_r^+ & \to G_{r +\eta}^+ \cr
 G_r^- & \to  G_{r-\eta}^- \cr}
}

The unitary representations were classified in \bfk. There is a
continuous series of unitary representations with $c\geq 3$ and a
discrete series with $c=3k/(k+2)$, $k\in \IZ_+$.

The $\NN=2$ discrete series can be realized by the GKO coset model
\eqn\supercoset{{SU(2)_k\times U(1)_2\over U(1)_{k+2}}.} The
representations are labelled by triples $(j,n,s)$
obtained by decomposing   $\CH_j^{SU(2)} \otimes \CH^{U(1)_2}_s$
with respect to a $U(1)_{k+2}$ subalgebra.
Thus representations are labelled by $(j,n,s)$ where  $j\in \{
0,\half, 1, \dots, k/2\}$, $n$ is an integer defined modulo $ 2k+4$,
and $s$ an integer defined modulo $4$.
There are four relevant $U(1)$ currents in the model
\supercoset. These are the $SU(2)$ current $J^3$,
the $U(1)_2$ current $J^F$, the $U(1)_{k+2}$ current
$J^{(k+2)}$ and the  $\NN=2$ current $J^{\NN=2}$. They
are related by
\eqn\relatecurrt{
\eqalign{
J^{\NN=2}& = - {2\over k+2} J^3 + {k\over 2(k+2)} J^F \cr
J^{(k+2)} & = 2 J^3 + J^F \cr}
}

If we express the four currents in terms of free scalar fields  as
\eqn\bosonize{
\eqalign{
J^3 & = i \sqrt{k\over 2} \p X \cr
J^F & = i 2 \p \sigma \cr
J^{N=2} &  = i \sqrt{k\over k+2} \p H \cr
J^{(k+2)} & = i \sqrt{2(k+2)} \p g \cr}
}
then
\eqn\relatebos{
\eqalign{
\sigma & = \sqrt{k\over k+2} H + \sqrt{2\over k+2} g \cr
X & = - \sqrt{2\over k+2} H  + \sqrt{k\over k+2} g \cr}
}
Using these we find
\eqn\decomp{
\Phi^{SU(2)}_{j,m} e^{i s \sigma/2} =
\biggl( \tilde \Phi
\exp\bigl[i ({s\over 2} - {(s+2m)\over (k+2)}) \sqrt{k+2\over k} H
\bigr] \biggr) \exp\bigl[ i {2m+s \over \sqrt{2(k+2)}} g\bigr]
}
Here $\Phi_{j,m}^{SU(2)}$ is any $SU(2)$ with those quantum
numbers, and $\tilde \Phi$ is a field in the super-parafermion theory.

{}From \decomp\  we conclude $n=(2m+s) \mod (2k+4)$, and hence the
selection rule
 \eqn\constraint{ 2j + n + s = 0 \mod 2 .}
The full set of equivalence relations on
the representation labels  $(j,n,s)$
is generated by
\eqn\equivrel{
\eqalign{
(j,n,s) & \sim (j, n+2k+4,s) \cr
(j,n,s) & \sim (j,n,s+4)\cr
(j,n,s) & \sim (k/2 - j , n+k+2, s+2 )\cr} }

{}From \decomp\ which one learns the $U(1)$ charge of the
representation $(j,n,s)$ is
\eqn\youone{  q_{(j,n,s)} =
\biggl[ {s \over 2} - {n\over k+2} \biggr] \mod 2
}
while the representations have $L_0$  values
\eqn\handpreq{
h_{(j,n,s)} = \biggl[ {j(j+1)\over k+2} -
 {n^2 \over 4 (k+2)} + {s^2\over 8} \biggr] \mod \IZ}
For the analysis of tachyons in the open string channels we need
more precise values for $h$. In the domain $\vert n-s\vert \leq
2j$ one may apply the expression in \handpreq\ provided one
minimizes over all values of $s\mod 4$. Explicitly we find:
\eqn\handq{ \eqalign{ h_{(j,n,s)} & = {j(j+1)\over k+2} - {n^2
\over 4(k+2)} + {s^2\over 8}\qquad -2j \leq (n-s) \leq 2j \cr
 & = {j(j+1)\over k+2} - {n^2 \over 4(k+2)} + {s^2\over 8}
+ {(n-s-2j )\over 2} \qquad 2j \leq (n-s) \leq 2k- 2j \cr}}
which is valid if we fix a fundamental domain $s=0,2,\pm 1$ for
$s$. Unfortunately, \handq\ does not cover the a full fundamental
domain for $n \mod (2k+4)$. Specifically, the exceptional cases
$(j,n,s)=(0,0,2), (0,-1,1),(0,1,-1)$ are not covered.  However,
by moving out of the fundamental domain of $s$ these can be
mapped to $({k\over 2},k+2,4), ({k\over 2},k+1,3), ({k\over 2},
-k - 1,-3)$ respectively.  Then the first expression in \handq\
can be used to find the dimensions $3\over 2$ and twice ${k\over
8(k+2)} +1={c\over 24} +1$.  In fact, by relaxing the domain of
$s$, it is enough to use only the first expression in \handq.  In
doing that we should minimize $h$ with respect to $s$ for fixed
value of $s\mod 4$.

There is an important operator of fermion number $(-1)^F$.
Choosing a fundamental domain it may be defined to be $(-1)^F=+1$
on representations with $s=0,1$ and $(-1)^F=-1$ on
representations with $s=2,3\sim -1$. The $\NN=2$ algebra mixes
these representations, but the chiral algebra does not. In
particular the only nonvanishing contributions to the Witten index
${\Tr}(-1)^F$ come from states $h_{(j,2j+1,+1)} =
h_{(j,-2j-1,-1)}= c/24$.

Under spectral flow by $\eta = +\half$ the representations
transform as $(j,n,s) \to (j,n+1,s+1)$, as one proves using the
$U(1)$ charge \youone.

The characters for the discrete series were found in
\refs{\gepner\ravanini-\qiu}
\eqn\characs{\eqalign{
 {\Tr}_{\CH_{(j,n,s)}} q^{L_0-c/24} e^{2\pi i z J_0} &=
 \chi_{(j,n,s)}(\tau,z) =\cr
 & \sum_{t=0}^{k-1} \chi^{(k)}_{j,n-s+4t}(\tau){1\over \eta(\tau)}
 \Theta_{2n+(4t-s)(k+2) ,2k(k+2)}(\tau, -{z\over 2k+4} ).}}

The modular transformation matrix is
\eqn\modtrmn{ S_{(j,n,s)}^{~~~(j',n',s')} = {1 \over k+2}
\sin\biggl( \pi {(2j+1)(2j'+1)\over k+2}\biggr) e^{ i \pi{
nn'\over (k+2)} } e^{- i \pi {ss'\over 2}} .}

The fusion rules are:
\eqn\fusionrules{ N_{jj'}^{~~j''} \delta_{n+n'-n''}
\delta_{s+s'-s''} + N_{jj'}^{~~{k\over 2} - j''}
\delta_{n+n'-(n''+k+2)} \delta_{s+s'-(s''+2)} }
where the delta functions are modulo $2k+4$ and $4$, respectively.

The discrete symmetry of the model $G$ can readily be found by
considering the coset $SU(2)_k\times U(1)_4\over U(1)_{2k+4}$. To
describe the symmetry group  let begin with the symmetry group of
the various factors in the coset $\IZ_2\times \IZ_4 \times
\IZ_{2k+4}$. Because of the spectral flow identification only a
subgroup of this group acts.  Moreover, due to the selection rule
$2j+n+s \in 2\IZ$ only a quotient of this subgroup acts
effectively. The resulting symmetry group $G$ is generated by
\eqn\discsymm{ \eqalign{ g_1 \cdot \Phi_{(j,n,s)} & = e^{2\pi i
\bigl({n\over 2k+4}- {s\over 4}\bigr) } \Phi_{(j,n,s)}\cr g_2
\cdot \Phi_{(j,n,s)} & = e^{i \pi  s} \Phi_{(j,n,s)}\cr} }
These generators satisfy the relations $g_1^{2k+4}=g_2^k$ and
$g_2^2=1$.  For $k $ even they generate $G=\IZ_{2k+4}\times \IZ_2$
and for $k$ odd the symmetry group is $G=\IZ_{4k+8}$. In the text
we usually restrict attention to the $H=\IZ_{k+2}\times \IZ_2$
subgroup generated by $g_1^2 g_2$ and $g_2$.


\appendix{F}{Simplest super-parafermion theory $k=1$}

The simplest nontrivial supersymmetric parafermion theory is based
on the coset
 \eqn\simplesp{{SU(2)_1\times U(1)_2\over U(1)_3}=U(1)_6.}
It has $c=1$ and representations:
 \eqn\sspararep{\matrix{ (n,s)&
\Delta & q\cr (0,0) & 0 & 0 \cr (\pm 2,0) & {2\over 3}& \mp
{2\over 3} \cr (\pm 2, 2) & {1\over 6}& \pm {1\over 3}\cr (0 , 2)
& {3\over 2}& 1\cr (\pm 1,\pm 1) & {1\over 24}& \pm {1\over 6} \cr
(\pm 1,\mp 1) & {25\over 24}& \mp {5\over 6}\cr (3 , \pm 1) &
 {3\over 8}& \mp {1\over 2}\cr}}
We used the identification $(j,s,n)\sim(\half-j,s+2\mod4, n+3 \mod
6)$ to set $j=0$ in all representations and did not write $j$.
$q=-{n\over 3}+{s\over 2} \mod 2$ is the $U(1)$ charge of the
representation.

We denote its chiral algebra by $\CS_2$.  It is the even fermion
number elements in the $\NN=2$ superconformal algebra. The
representations with $s=0\mod 2$ arise from the NS sector. They
are combined to the three NS $\NN=2$ representations with
$(\Delta,q)=(0,0),({1\over 6},\pm {1\over 3})$. The
representations with $s=1\mod 2$ arise from the R sector. They
combine into the three R $\NN=2$ representations with
$(\Delta,q)=({1\over 24},\pm {1\over 6}),({3\over 8},{1\over 2})$.

Since $c$ is less than ${3\over 2}$, this theory is also a member
of the $N=1$ discrete series and hence can be represented as the
coset $SU(2)_2\times SU(2)_2 \over SU(2)_4$ (note that the chiral
algebra of this coset, which we denote as
$\CS_1$, is only the even fermion number
elements of the $N=1$ superconformal algebra).  The
representations of this coset are labeled by the three spins
$(j_1,j_2,j_3)$ subject to the selection rule $j_1+j_2+j_3\in
\IZ$ and the identification $(j_1,j_2,j_3)\sim
(1-j_1,1-j_2,2-j_3)$.  As always, there are two distinct
representations at the fixed point of the identification
$(j_1={1\over 2},j_2={1\over 2},j_3=1)_\pm$.  The complete list
of representations of the coset is
 \eqn\nonecoset{\matrix{
(j_1,j_2,j_3)& \Delta \cr (0,0,0) & 0 \cr (0,{1\over 2},{1\over
2}) & {1\over 16} \cr (0,1, 1) & {1\over 6}& \cr (1,{1\over
2},{1\over 2}) & {9\over 16}\cr (0,0,1) & {2\over 3} \cr (1,1,0)
& 1\cr (0,1 , 0) & {3\over 2}\cr (1,0 , 0) & {3\over 2}\cr
({1\over 2},{1\over 2},1) & {1\over 24} \cr ({1\over 2},{1\over
2},1)' & {25\over 24} \cr ({1\over 2},0,{1\over 2}) & {1\over 16}
\cr ({1\over 2},{1\over 2},0) & {3\over 8} \cr ({1\over
 2},1,{1\over 2}) & {9\over 16} \cr }}
The representations with $j_1\in \IZ$ are in the NS sector and
those with $j_1\in \IZ + \half$ are in the R sector.  The
representations of the $N=1$ superconformal algebra are
 \eqn\nonealge{\matrix{(j_1,j_2,j_3)&\Delta& (r,s)\cr
(0,0,0)\oplus (1,0,0) & 0 &(1,1) \cr (0,{1\over 2},{1\over
2})\oplus (1,{1\over 2},{1\over 2}) & {1\over 16} & (2,2) \cr
(0,1,1) \oplus (0,0,1) & {1\over 6} & (1,3) \cr (1,1,0) \oplus
(0,1,0)& 1 & (1,5) \cr ({1\over 2},{1\over 2},1) \oplus ({1\over
2},{1\over 2},1)' & {1\over 24} & (2,3) \cr 2({1\over
2},0,{1\over 2}) & {1\over 16} & (1,2) \cr 2({1\over 2},{1\over
2},0) & {3\over 8} & (2,1)\cr 2({1\over 2},1,{1\over 2}) &
 {9\over 16} & (1,3) \cr }}
In the last column we included the $(r,s)$ values of the
representation in the Kac table.

The super-parafermion theory is not the standard modular
invariant of this theory but the one with an extended chiral
algebra. It can be denoted as $SU(2)_2\times SU(2)_2 \over
SO(3)_4$.  Its representations are obtained from the original
coset by imposing the selection rule $j_3\in \IZ$, the
identification $(j_1,j_2,j_3)\sim (j_1,j_2,2-j_3)$ and doubling
the representations at the fixed points $(j_1,j_2,j_3=1)$ and
$(\half,\half,0)$.  This leads to the same spectrum as \sspararep\
\eqn\ssspararep{\matrix{ (n,s)& (j_1,j_2,j_3)&\Delta & q\cr (0,0)
&(0,0,0)\oplus (1,1,0)& 0 & 0 \cr (\pm 2,0) & (0,0,1)&{2\over 3}&
\mp {2\over 3} \cr (\pm 2, 2)& (0,1,1) & {1\over 6}& \pm {1\over
3}\cr (0, 2) &(0,1,0)\oplus (1,0,0)& {3\over 2}& 1\cr (\pm 1,\pm
1)&({1\over 2},{1\over 2},1)_+&{1\over 24}& \pm {1\over 6}\cr
(\pm 1,\mp 1)&({1\over 2},{1\over 2},1)_- & {25\over 24}& \mp
{5\over 6}\cr (3,\pm 1)& ({1\over 2},{1\over 2},0) & {3\over 8}&
\mp {1\over 2}\cr}}

The partition function of the super-parafermion theory is
\eqn\paraparti{Z_{PF}=|\chi_{0}+\chi_{1}|^2+2|\chi_{1\over 6}|^2
+2|\chi_{2\over 3}|^2+|\chi_{3\over 2}+\chi_{{3\over 2}'}|^2
+2|\chi_{1\over 24}|^2+2|\chi_{25\over 24}|^2 +2|\chi_{3\over
8}|^2,} where we have labeled the $\CS_1$ characters by the
dimension of the primary.  The character $\chi_1 $ which extends
the chiral algebra includes the $U(1)$ current.  The two
dimension $3\over 2$ fields which are primary fields of this
algebra are the two supercharges of the $\NN=2$ algebra. We also
recognize the two R ground states with $U(1)$ charges $\pm
{1\over 6}$.

This theory clearly has 12 Ishibashi states $|Ans\rangle\rangle$
for $ns$ in \sspararep, which are invariant under the full chiral
algebra of the theory $\CS_2$.  Using these states and the
S-matrix
 \eqn\superparas{S_{ns}^{~~n's'}= {1 \over
 \sqrt{12}}e^{{i\pi nn'\over 3} - {i\pi ss' \over 2}}}
we can construct 12 Cardy states
 \eqn\superparca{|A,\hat n \hat s\rangle_C=\sum_{\hat n'\hat s'}
 {S_{\hat n\hat s}^{~~ n's'}\over \sqrt{S_{00}^{~~n's'}}}|A, n's'
 \rangle\rangle.}
(These are exactly the 12 A Cardy states of the $U(1)_6$ theory.)

Using our standard picture we interpret the target space as a
disk with 6 marked points on its boundary.  The 12 Cardy states
correspond to 12 D-brane states.  These are oriented D1-branes
stretched between the even or odd marked points.

There are also other branes which respect only $\NN=1$
supersymmetry.  These are the B-branes which turn out to preserve
another $\NN=2$ supersymmetry.  We start by looking for Ishibashi
states preserving $\CS_1$.  Since the boundary state has to
create closed strings which exist in the spectrum, we look at
\ssspararep\paraparti\ and consider the Ishibashi states $|
0,0,0\rangle\rangle-| 0,0,2\rangle\rangle$ and $| 0,1,0)
\rangle\rangle-|1,0,0\rangle\rangle$ (we use the notation in
terms of $(j_1,j_2,j_3)$ because these are not Ishibashi states
of $\CS_2$ and therefore cannot be labeled by $ns$).  We define
the two Cardy states labeled by $\hat s=0,1$
 \eqn\dzeroca{|B,\hat s\rangle_C=3^{1\over 4} \left[|0,0,0\rangle
 \rangle-|0,0,2\rangle\rangle +(-1)^{\hat s}(|1,0,0\rangle\rangle
 -|0,1,0\rangle\rangle)\right].}
(These are exactly the two B Cardy states of the $U(1)_6$ theory.)
These two states are even and odd Cardy states of the $\CS_1$
algebra. They are annihilated by $G(\sigma)-i(-1)^{\hat s}\bar
G(\sigma)$.   We interpret them as a single D0-brane at the
center of our disk or a D1-brane around the $U(1)_2$ circle.

It is easy to compute the inner product
 \eqn\dzdzinns{\eqalign{
 {}_C\langle B,\hat s|&q_c^{L_0-c/24}|B,\hat s'\rangle_C=\cr
 &\sqrt 3(\chi_{(0,0,0)} (q_c) +\chi_{(0,0,2)}(q_c)+ (-1)^{\hat s
 +\hat s'} (\chi_{(0,1,0)}(q_c)+ \chi_{(1,0,0)}(q_c))=\cr
 &\sqrt 3(\chi_{(0,0)}(q_c)+(-1)^{\hat s+\hat s'} \chi_{(0,2)}
 (q_c))= \sum_{ms''}{1+(-1)^{\hat s+\hat s'+s''}\over 2}
 \chi_{(m,s'')}(q_o).}}
Here we first expressed the answer in terms of $\CS_1$
characters, then in terms of $\CS_2$ characters, and finally in
terms of characters in the open string channel with modular
parameter $q_o$.

We see that the open strings living on these two D-branes are the
same and include all the NS representations of $\CS_2$.  The
spectrum of string stretched between them is the R representations
of $\CS_2$.

Using the S-matrix of the $N=1$ models it is easy to compute the
inner products
 \eqn\dzdoinns{\eqalign{
 {}_C\langle &B,\hat s'|q_c^{L_0-c/24}|A,\hat m, \hat s\rangle_C=\cr
 &{1\over \sqrt 2}(\chi_{(0,0,0)}(q_c)-\chi_{(0,0,2)}(q_c)+
 (-1)^{\hat s+\hat s'}(\chi_{(0,1,0)}(q_c)-\chi_{(1,0,0)}(q_c)))=\cr
 &{1+(-1)^{\hat s+\hat s'}\over 2}(\chi_{(0,{1\over 2},{1\over 2})}
 (q_o)+\chi_{(1,{1\over 2},{1\over 2})}(q_o))+ {1-(-1)^{\hat s+\hat
 s'}\over 2} (\chi_{({1\over 2},0,{1\over 2})}(q_o)+\chi_{({1\over
 2},1,{1 \over 2})}(q_o)),}}
where all these characters are $\CS_1$ characters which are
labeled by $(j_1,j_2,j_3)$. We see that we get only $\NN=1$
representations which are not $\NN=2$ representations.  Also, for
$\hat s=\hat s'$ we get only the representations in the NS sector
and for $\hat s\not=\hat s'$ only the representations in the R
sector.

To summarize, both the A-branes and the B-branes preserve $\NN=2$
supersymmetry. Therefore, the open strings stretched between two
A-branes or between two B-branes are in $\NN=2$ representations.
However, since these two kinds of branes preserve different
$\NN=2$ symmetries, the strings stretched between A-branes and
B-branes are not in $\NN=2$ representations and are only in
$\NN=1$ representations.

The special point about this example is that since this model is
a member of the $\NN=1$ discrete series, we know that there are
no other branes which preserve $\NN=1$ supersymmetry.  There
might be, however, other branes which preserve only the Virasoro
algebra and which satisfy all other consistency conditions.

\listrefs
\end